\documentclass[12pt, a4paper]{article}
\usepackage[pdftex]{graphicx}

\usepackage{hyperref}
\usepackage{algorithm}
\usepackage{algorithmic}
\usepackage[utf8]{inputenc}
\usepackage{subfigure}
\usepackage{csquotes}
\usepackage{placeins}
\usepackage{color}
\usepackage{listings}
\usepackage{authblk}
\usepackage[official]{eurosym}

\DeclareFontFamily{OT1}{cmtt}{\hyphenchar \font=-1}
\DeclareFontFamily{\encodingdefault}{\ttdefault}{\hyphenchar\font=`\-}
\DeclareFontFamily{T1}{cmtt}{\hyphenchar \font=45}

\definecolor{terminalColor}{RGB}{245,245,245}
\definecolor{terminalFont}{RGB}{38,50,56}
\lstdefinestyle{myterm}{
    basicstyle=\color{terminalFont}\ttfamily\scriptsize,
    backgroundcolor=\color{terminalColor},
    commentstyle=\color{codegreen},
    keywordstyle=\color{magenta},
    stringstyle=\color{codepurple},
    breakatwhitespace=false,
    breaklines=true,
    captionpos=b,
    keepspaces=true,
    showspaces=false,
    showstringspaces=false,
    showtabs=false,
    tabsize=2,
    frame=shadowbox,
    rulecolor=\color{white},
}

\lstdefinestyle{mycode}{
    language=C,
    breaklines=true,
    basicstyle=\tt\scriptsize,
    keywordstyle=\color{blue},
    identifierstyle=\color{magenta},
    frame = single
}

\begin{document}
\title{Auditable and Transparent Fully Authenticated Disk Encryption via USB Storage Interposition}

\author[1]{Enrique Soriano-Salvador\footnote{Corresponding author: enrique.soriano@urjc.es}}
\author[1]{Gorka Guardiola Múzquiz}
\affil[1]{Universidad Rey Juan Carlos, Madrid, Spain}

\maketitle
\begin{abstract}
Full Disk Encryption (FDE) has become increasingly important in the
last decades due to the evident confidentiality concerns. In most systems,
encryption is provided by an operating system driver, through which the
user can transparently access the encrypted disk after supplying the
required keys (or the credentials from which those keys are derived). In
this work, we explore an alternative approach: the use of an intermediate
USB device placed between the host system and the external hard disk,
where the encrypted data are stored.  Although there are existing
devices that follow this \emph{inline} approach (such as RAID controllers
and USB enclosures with built-in encryption),
we explore the use of a general-purpose single-board computer
running Linux with USB On-The-Go support (e.g. a Raspberry Pi), to
provide FADE (Fully Authenticated Disk Encryption).
Our system, named CC (\emph{Cryptographic
Companion}), performs the cryptographic operations (encryption/decryption
and authentication), providing
a standard USB mass-storage interface to the host (which is entirely
unaware of the presence of encryption) while relying on the external
USB hard disk to store the corresponding encrypted blocks.  Our design
provides several key advantages: flexibility, low cost, transparency,
the use of generic hardware and free and open-source software,
adaptability to emerging cryptographic schemes, and mitigations against
malicious disk firmware. This paper
presents the design and implementation of the CC and an experimental
evaluation of our current research prototype, which indicates that it
is sufficiently efficient for most common use cases.
\end{abstract}

\section{Introduction}

Full Disk Encryption (FDE) has become an essential mechanism
for protecting data \emph{at-rest}, particularly in response to growing
confidentiality concerns arising from device loss, theft, or unauthorized
physical access.

Commonly, FDE is implemented within the operating system, typically
through a dedicated storage or block device driver. This approach
allows encryption and decryption to be performed transparently:
applications and users interact with the disk in the usual manner
once the required cryptographic keys (or credentials from which those
keys are derived) have been provided. While this design integrates
naturally with existing software stacks, it tightly couples the
encryption mechanism to the host operating system,
which may limit flexibility and portability across
heterogeneous environments.

In this work, we explore an alternative approach to provide FDE with
full authentication, or FADE
(Fully Authenticated Disk Encryption)~\cite{BENADJILA2022100465},
based on the use of an intermediate USB device placed between
the host system and an external hard disk. In our model, encrypted
data are stored exclusively on the external disk, while the intermediate
device transparently handles encryption and decryption operations. Similar
approaches are already employed by using specialized hardware, such
as RAID controllers or disk enclosures
with built-in cryptographic capabilities.
However, these
solutions are typically closed, non-auditable, inflexible, and tied
to specific hardware implementations.

Instead, we explore the use of a general-purpose
single-board computer (from now on, SBC) running Linux with USB On-The-Go support
(for example, s a Raspberry Pi or an Orange Pi) to realize this functionality. Our
system, named CC (\emph{Cryptographic Companion}), acts as a cryptographic
mediator between the host and the storage device.
The intended design envisions the CC as a compact USB dongle with two connectors,
functioning as an inline adapter between the external disk and the host.
Ongoing advances in hardware miniaturization suggest that a device with
this form factor is achievable (see for example the Raspberry Pi Zero series
or the Allwinner SoC family).

The CC presents a standard USB mass-storage interface to the host, exposing
only decrypted data blocks. Thus, the host is entirely unaware of
the presence of encryption. At the same time, the external USB hard disk
connected to the CC is used to store the corresponding encrypted blocks. This
design cleanly separates cryptographic functionality from the host system,
enabling encryption to be enforced independently of the host operating
system. This way, any legacy system with USB mass storage support is
able to use full disk encryption without any new specialized driver
or software.

Our approach offers several key advantages. By relying on generic
hardware and free/libre software, the system remains
low-cost, auditable\footnote{The source code of the whole software stack
is available for reviewing and it is posible to recreate and install any of the components if necessary.},
and easily reproducible. Moreover, its software based
nature provides significant flexibility, allowing the cryptographic
implementation to be modified or extended without changes to the host
or storage hardware. In particular, the design facilitates adaptation to
emerging cryptographic schemes.

An additional key advantage of the approach is that it can eliminate
the need to store encrypted keys or any other cryptographic metadata in the
external disk headers, a practice commonly adopted by software-based
FDE systems. As a result, the external disk stores exclusively encrypted
data blocks and contains no cryptographic material that could aid an
attacker in cryptanalytic efforts: storage contents
exhibit uniformly high entropy across the entire disk.

In addition, our CC implementation includes mitigations against malicious
disk firmware, including disk block address remapping and deception
techniques designed to hinder the detection of specific disk blocks
storing known data, thereby mitigating known-plaintext cryptanalysis.

\subsection{Contributions}

In short, the contributions of this paper are:

\begin{itemize}
	  \item We present the design and the architecture of
	  the \emph{Crypto Companion},
	  a general-purpose, software-based FADE (Fully Authenticated
	  Disk Encryption) interposed system based on commodity hardware
	  and free/libre software.
%%V2
	  To the best of our knowledge, no previous work has followed the
	  same approach.

	  \item We describe the implementation of our functional research
	  prototype written in Go, based on common and widely adopted
	  cryptographic algorithms (AES-CTR and HMAC-SHA1).
%%V2
	  \item We provide an experimental evaluation of the prototype
	  running on an Orange Pi 5 Ultra
	  SBC\footnote{Note that this hardware is selected solely for research
	  prototyping and evaluating the proposed approach, and is not intended as a platform for
	  developing a final product.},
	  showing the costs of the approach.
	  Two different standard I/O benchmarks (FIO~\cite{fio} and
	  Filebench~\cite{tarasov2016filebench}) have been used to conduit
	  different experiments. The results suggest that
	  the proposed design achieves sufficient performance for most common
	  use cases, while preserving transparency and flexibility.
\end{itemize}

\subsection{Organization}

The rest of the paper is organized as follows:
Section \ref{related} presents the related work;
Section \ref{model} explains our threat model;
Section \ref{arch} describes the architecture of the system;
Section \ref{impl} describes the current research prototype implementation;
Section \ref{eval} explains the evaluation of the prototype;
Section \ref{limitations} presents some limitations of the present work;
finally, Section \ref{conclusions} presents the conclusions.

\section{Related Work \label{related}}

To the best of our knowledge, no prior work has proposed a
system analogous to the scheme presented in this paper, based on the use of an
SBC to perform encryption transparently between the host
machine and a conventional storage device. In general, few approaches
have leveraged the capabilities of USB On-The-Go (OTG), as supported by
modern SBCs, to enhance storage security. In earlier work,
we presented a different system built upon such platform to provide write-once
read-many (WORM) storage devices~\cite{soca}; however, no other comparable solutions
have been identified.

\subsection{Full Disk Encryption Software}
%% https://en.wikipedia.org/wiki/Comparison_of_disk_encryption_software
%% https://en.wikipedia.org/wiki/Disk_encryption_theory

Although alternative paradigms for storage encryption exist (e.g. file system-level
encryption~\cite{Baodian,halcrow2005ecryptfs} or deniable
file systems~\cite{czeskis2008defeating,10.1007/BFb0052229,durmuth2011deniable,howlader2009sender,zhang2018ensuring}),
our system is directly related to full-disk encryption (FDE).
There is a wide range of alternative approaches for providing
FDE\footnote{
These systems can be classified according to whether
they provide full-disk encryption (including partition tables, among
other components), whether they support the use of a simple
file to host an encrypted volume (e.g., via a loop device), and
related design characteristics. See for example \url{https://en.wikipedia.org/wiki/Comparison_of_disk_encryption_software}.}.
At present, the most popular
conventional full disk encryption systems are
BitLocker~\cite{bitlocker},
FileVault~\cite{filevault},
dm-crypt/LUKS (Linux Unified Key Setup)~\cite{dmcrypt,8678978,luksusers} or
VeraCrypt~\cite{veracrypt}. These systems are widely adopted.

M{\"u}ller~\cite{6951337} et al. presented an overview
of software-based and hardware-based FDE.
Later, Benadjila et al.~\cite{BENADJILA2022100465}
published a study of FDE solutions and the related threat models.
These works can provide a comprehensive survey of the state of the art
on FDE. We will focus on the differences between our system and
existing approaches.

In addition to encryption, the FDE system can provide authentication
for the encrypted data. In this case, it can be
referred to as ADE (Authenticated Disk
Encryption)~\cite{BENADJILA2022100465,khati2019full}.
In those systems (see for example~\cite{10.1007/978-3-319-99828-2_6}),
if a disk block is modified directly by the attacker,
it will be detected by the system.
However, under ADE, an attacker
may still be able to replace a block with a previous version of the
same block (that is, perform a replay attack on that block)~\cite{ade}.
In contrast,
such attacks are not possible in
FADE (Fully Authenticated Disk Encryption)
systems~\cite{7345286,10.1145/3124680.3124732,10.1145/3173162.3173183}.
Our system provides FADE in a very simple way, as we will discuss later.

FDE is usually implemented as an operating system
driver and low level software. For example, LUKS is implemented
over another component, \texttt{dm-crypt}~\cite{dmcrypt},
as the disk encryption backend. It follows a
layered design split between user-space management and
kernel-space for device mapping and data encryption/decryption.
Thus, these drivers and user-space programs have to be ported
to other operating systems.
For example, an old system running Microsoft Windows 7
or some families of newer versions (such as \emph{home editions})
are not able to use a disk encrypted with BitLocker.
On the other hand, any operating system
that supports USB disks, including embedded and plropietary systems,
is able to use the CC.

Such systems may employ a variety of cryptographic schemes specifically designed
for \emph{at-rest} encryption.
Specialized hardware (addressed in the next subsection)
also use these algorithms.
A fundamental design principle of these schemes is that the encryption
method should not incur any disk space overhead.
For instance, they avoid the use of initialization vectors (IV) or \emph{nonces}
to encrypt the disk blocks,
that would be stored in the disk itself (reducing the total
storage capacity).
This property entails inherent trade-offs and cannot be achieved without cost.

Some schemes rely on standard modes of operation combined with specific methods
for generating initialization vectors, such as the
Encrypted Salt-Sector Initialization Vector (ESSIV).
GEOM~\cite{geom} proposed an approach in which a
distinct encryption key is generated for each disk block.
Liskov et al.~\cite{liskov2002tweakable} presented the idea of \emph{tweakable}
ciphers (LWR encryption).
Another tweakable encryption mode, XEX, was formally introduced
by Rogaway~\cite{xex}.
Certain security concerns were identified with LRW
mode of operation, so it was superseded by XTS (XEX-based tweaked-codebook
mode with ciphertext stealing), which became a standard~\cite{xtsstandard}.
Nevertheless, XTS is susceptible to data manipulation
and tampering. Applications must employ extra measures to detect
modifications of the data if manipulation and tampering is a concern.
Moreover, as the ECB mode, XTS fails to hide patterns: ECB shows
patterns in space, while XTS shows patterns over time (i.e. in different
versions of the same encrypted block).
Adiantum~\cite{Crowley_Biggers_2018} is an alternative mode to XTS,
originally designed by Google for deployment in Android systems. It
is optimized for performance on resource-constrained devices lacking
hardware acceleration for AES.
StrongBox~\cite{10.1145/3173162.3173183} proposed the use of
stream ciphers for FDE, achieving good performance on ARM devices.
Secure Block Device~\cite{7345286} proposed an approach
based on Merkle-Trees in conjunction with a selectable
authenticated encryption scheme.

The CC can employ any encryption algorithm or mode without
incurring any disk space overhead,
because it can use its own alternative storage
(e.g. the internal volume of the SBC
or a separated SD Card or USB drive) to store
the needed metadata, for one or several ciphered disks. For example,
our current prototype uses common cryptographic algorithms
(AES-CTR and HMAC-SHA1).

Moreover, FDE software usually rely on encrypted keys and other
cryptographic metadata stored in some parts of the encrypted
disk (i.e. the headers). For example, LUKS uses several headers to
store metadata.
%%V2
Such systems are generally considered secure, provided that they are
properly designed and implemented.
Nevertheless, even when metadata and keys are stored securely, they may still aid
cryptanalysis of the encrypted blocks.
For example, if the encryption
keys are stored encrypted with a key derived from the user password,
the attacker may be able to break the password, specially if it is
weak (e.g. by performing offline attacks, etc.).
In our approach, the encryption metadata can be stored in the CC itself.
This way, the encrypted disk only stores encrypted data,
thereby increasing the effective security of the disk. Offline password attacks
are not viable without the CC.

With respect to cold boot-style
attacks~\cite{coldboot,amnesia,huber2016flexible}, the CC is as
vulnerable as any conventional software-based FDE, since the keys reside in
memory (in our case, in the SBC's memory). Note that,
like some software FDE systems, the CC could take advantage of security
hardware (e.g. TPM or secure enclaves) when available in the SBC.

\subsection{Specialized Hardware}

HDDs or SSDs with built-in full-disk encryption, or Self-encrypting drives (SED)
(e.g. those following the Opal~\cite{opal,opalcore} and
IEEE 1667~\cite{8479380} specifications)
are an alternative to FDE software.
SATA/NVMe to USB bridges and RAID controllers
can also provide builtin encryption, see for example
the maxCrypto RAID controllers of Microchip Technology Inc.
SEDs, bridges and RAID controllers le encryption inside the hardware instead
of relying on operating system or low level software. They are transparent
for the user and provide good performance. Moreover, they are
independent of the operating system (as our approach).

Hardware-based solutions offer certain advantages; however, the hardware
itself may also be subject to attacks.
For example, Skorobogatov~\cite{ironkey} exposed some flaws in the IronKey
USB drive (a FIPS 140-2 Level 3 device).
Moreover, hardware solutions are closed and only the vendor is able to patch
the vulnerabilities of the firmware.
Meijer et al.~\cite{8835339} analyzed the firmware of several
SED models by reverse engineering. For many models, they recovered
the cleartext data without knowledge of any key or password.
Muller et al.~\cite{muller2012self} also analyzed the security of SED
and were able to compromise the security of several models.

Some commercial external encrypted disks, such as the Apricorn Aegis or
the iStorage diskAshur, have a physical numerical keyboard
to provide a PIN to encrypt the data. A PIN lacks sufficient entropy for the
generation of strong keys.
Other drives, like the Viasat Eclypt Freedom, can use an external
USB key to provide multi-factor authentication.
Note that the CC can implement different schemes for user authentication
and key derivation, because it is a complete Linux system.
Some of these controllers can take advantage of the TPM hardware.
Our approach could use security hardware as well.

The product that most closely resembles our approach is
the disk enclosures that provide encryption. For example,
the ICY BOX IB-246FP-C3 is an external
USB enclosures with a fingerprint reader that provides AES encryption.
Note that this device depends on a proprietary OS to
configure the biometric information.
Another example is the StarTech S251 Enclosure.
Our goal is to implement a cost-effective open and auditable
device based on a general purpose SBCs and common USB disks,
allowing compatibility with different conventional hard drives
(i.e. one CC can be used with several common USB disks),

The security of closed hardware depends entirely on their
firmware and hardware design. Thus, the vendor must be trusted.
Avoiding dependence on specialized and proprietary
hardware and software makes the CC auditable and modifiable.

In our scenario and in the software FDE scenario,
trust in the disk's vendor is not required, because
the disk only handles encrypted data.
In addition, we implement different mitigations
against malicious disk firmware.
We will address this issue in
Section~\ref{model}.

Purely hardware based solutions present a clear drawback.
Besides relying on encryption algorithms originally
designed for data storage (like software FDE), which
have known weaknesses (such as XTS), they lack the flexibility
to transition to new encryption schemes that may be required
in the future, since only the vendor can update the firmware.

With respect to hot plug attacks~\cite{muller2012self},  the CC is as vulnerable
as SED: if an attacker is able to switch the USB cable to connect a different host
without interrupting the operation of the cryptographic controller (CC), she
may gain access to the disk blocks in plaintext (as the CC is indistinguishable
from an unencrypted disk from the host’s perspective).

\subsection{Network Storage}

Diffferent researchers have been proposing network-based schemes for encrypted
storage for a long time.
For example, Miller et al.~\cite{miller2002strong}
introduced a cryptographic system designed to meet the security and performance
requirements of network-attached storage (NAS) systems, separating storage management
from the underlying storage device.
Aguilera et al.~\cite{nadsec}  also proposed a security scheme
for network-attached systems and storage management
software. They added minor changes to the standard
protocol to provide security and required no modification to the disks.
Zhu et al.~\cite{snare} presented SNARE, a security scheme for NAS
based on a capability-based model that uses a key distribution mechanism
that eliminates the need for the storage system itself to perform key
management.
Another example is Guardat~\cite{guardat}, a system that enforces
file protection policies (expressed in a custom language) in the storage layer
in storage area network (SAN) servers (in this case, iSCSI).

Nowadays, there are a plethora of NAS and SAN
solutions for consumer, business, enterprise and defense environments that offer a
disk encryption features. Ultimately, the former are equivalent
to deploying a RAID controller with encryption capabilities within a network server
or use SED.
SAN solutions generally offload encryption to SED or software FDE in the client side.
In this work, we focus on local storage rather than on network-based storage systems.

\section{Threat Model \label{model}}

The model essentially corresponds to the one applied to FDE,
guaranteeing the three key properties of disk encryption:

\begin{enumerate}
	\item  The information stored on the disk's blocks must remain confidential.
	In addition, we provide authentication for the data following the
	FADE model.

	\item Reading from and writing to the disk should be efficient operations.

	\item The system should utilize storage space efficiently,
	ensuring that the size of the
	encrypted data does not significantly exceed that of the original plaintext.
	In our case, the disk size of the encrypted disk is the same as the plaintext
	disk (i.e. the encryption is totally transparent for the host).
\end{enumerate}

%%V2
A compromise of the CC results in access to the cryptographic keys
and metadata, as well as to the information exchanged between the
disk and the host, both in encrypted form and in plaintext.

\subsection{Threats}

We consider that the attacker is able to
read and write the raw contents of the disk at any moment,
with multiple passive or active accesses to the disk (e.g. to
create copies of the disk to track modifications between versions,
tamper with the disk data, etc.), and perform online (accessing
to the disk hardware or firmware) and offline attacks.

%%V2
The host is trusted by assumption. The purpose of our system is
to provide confidentiality and authenticity (i.e. \emph{at-rest} encryption),
for data stored on an external drive.
Once the CC has been initialized correctly with the corresponding keys\footnote{Note
that there are different methods for key management, see Section \ref{keymanagement}.}
and enters operational mode,
it serves the contents of the external disk
in plaintext to the host. Consequently, the host system can access and modify
the data without restriction. In such a scenario, considering a malicious
host is therefore outside the scope of the threat model.

%%V2
The host runs a conventional operating system and can rely on standard
access control mechanisms to manage which users are allowed to access the volume exposed
by the CC.  For example, the operating system can control which users are
allowed to mount the volume and which users have read or write access to
the mounted file system. Thus, malicious users on the host may be considered
within the model, with protection against them being analogous to
that provided by conventional full-disk encryption systems (FDE or SED).
However, if the host system itself is compromised or malicious (e.g.
if the kernel of the host system is compromised or any malware is running
with sufficient privileges), it can access and modify the data
served by the CC. This limitation is inherent to the considered architecture
and is equivalent to that of conventional full-disk encryption systems.

%%V2
Interaction with the CC by an adversary is restricted exclusively
to the USB interface connected to the disk.
As explained in the introduction, we envision the CC as a dedicated
device with two USB ports, one for connecting to the host and the
other to the external drive, running our system as a complete
software image that includes both the operating system and our
software, all released as open-source/libre software (to be auditable).
For a conventional SBC acting as the CC (like in our research prototype),
we assume that the device is
flashed with our complete system image, including the CC software,
has all wired and wireless network connectivity disabled,
runs exclusively the CC services, and has no external devices
connected during operation.

We also consider that the attacker can analyze
the block usage (the addresses)
and record the operation history, in order
to detect patterns and find blocks
that are valuable to conduit a cryptoanalysis.
This could be performed by malicious disk firmware.

\subsection{Dependencies and Assumptions}

The hardware platform for CC devices is assumed
to be conventional, for example, an ARM-based SBC equipped with standard
storage components such as SSD, SD, or NVMe devices.
Both the hardware and software components of the CC are assumed to
function correctly and remain trustworthy.
%%V2
This assumption requires the CC system to boot from storage protected by
a secure root of trust that guarantees the integrity and
authenticity of the system components.

The CC includes internal storage with sufficient capacity to hold
the metadata required for disk encryption and authentication,
as well as the associated configuration parameters.

The host may operate in either online or offline mode.
The CC device is connected to the host via a USB interface.
The integrity of this link is assumed to be preserved and it operates
strictly within the specifications of a standard mass storage device.

To summarize, the attack surface of the CC corresponds
to the two interfaces, one in which it behaves as
a USB disk reading and writing plaintext blocks (towards the host)
and another in which it is reading and writing encrypted blocks
to an external disk. We assume the integrity of both.

\subsection{Mitigation \label{threatmitigation}}

When the data of a block is not correctly authenticated, the CC will provide
I/O errors for the corresponding disk operation. For blocks that have
never been written by the host system,
the host always reads zeros (independently of the data stored in the disk).

%%V2
In addition, the CC includes two mechanisms designed to hinder the identification
of the addresses of blocks containing sensitive encrypted content.
For example, when a disk is connected, the block with address 0 is
read by the operating system to detect the partitions.
Also, when a file system is mounted, the superblock and other parts
of the metadata are read. Those blocks may store encrypted blocks
with a fully or partially known plaintext.
Therefore, they may constitute valuable assets for cryptanalytic purposes.
The following two \emph{mitigations}\footnote{In this case, we use the
term \emph{mitigation} in its strict sense: an action
taken to reduce or limit the damage or impact of a vulnerability (in this case,
the identification of sensitive encrypted blocks through their addresses).
That is, we are not
referring to security defenses intended to prevent or completely
block an attack.}
are transparent for the host system:

\begin{itemize}
	\item Random mapping. The physical addresses of the
	device provided by the CC to the host are not the physical addresses
	of the disk. The CC can apply a key based permutation to map them.
	Different algorithms can be used for this permutation.

	\item Ghost reads. When a sensitive read-only
	block is read by the host, the CC can
	perform \emph{fake} read operations for a deterministic random set of blocks
	(i.e. always the same set, in the same order, for a target address).
	Ghost reads are
	discarded by the CC: only the correct plaintext block is provided
	to the host.
	The addresses of the blocks with this protection and the size of
	the set is defined by the user in the configuration.
\end{itemize}

These mitigations will inherently introduce performance
overhead on disk operations.
The user should be granted the ability to disable both options
through the configuration settings.

\section{Architecture \label{arch}}

\begin{figure}[th]
\begin{center}

	\resizebox{\columnwidth}{!}{
		\includegraphics{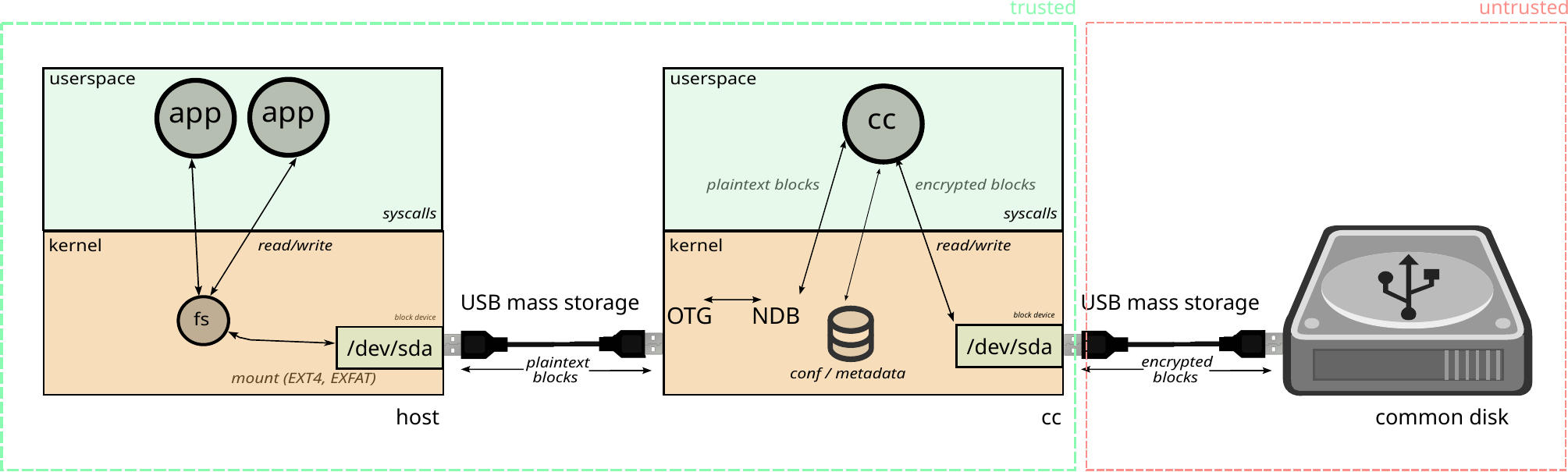}
	}
	\caption{Architecture of the system. \label{figarch}}
\end{center}
\end{figure}

The general architecture of the system is depicted in Figure~\ref{figarch}.

%%V2
The SBC is connected to the host when it starts (it can be powered by
the USB cable).  Then, a regular untrusted USB drive is connected to the SBC,
so the Linux kernel offers it as a standard disk device (for example,
\texttt{/dev/sdb}). Then, a service (a \texttt{systemd} unit) running
in the SBC detects the disk and starts the CC component, which is a
userspace program.  The CC finds the ID of the disk and retrieves its
configuration (parameters and the metadata required to decrypt, encrypt
and authenticate the blocks), which are stored in the SBC’s internal
storage (or in other external drive, such as a SD card).

The CC reads and writes encrypted blocks by accessing \texttt{/dev/sdb}
with the standard system calls (\texttt{read} and \texttt{write}), and
serves decrypted blocks through OTG. To serve the disk blocks to the
host, it uses the Network Block Device (NBD) protocol~\cite{nbd} over
an internal Unix domain socket\footnote{NBD consists of two interfaces:
a network client-server protocol to export block devices, and an operating
system interface that allows a user-space client (nbd-client) to map the
remote device as a regular block device. In this architecture, both the client
and server run on the CC and communicate via a Unix domain socket. More
details about NBD can be found in our previous work~\cite{soca}.}.

After initialization, the CC exports the NBD device (typically
\texttt{/dev/nbd0}) via USB using OTG.  From the host’s perspective,
the CC appears as a regular block device, mapped to a device node under
\texttt{/dev/sda}.  This block device is mounted by the host's operating
system, allowing applications to access and use it transparently as with
any disk.

When the user finishes using the disk, the host operating system unmounts
the file system. Subsequently, the user presses a button on the SBC,
signaling the service to terminate the operation associated with the
connected disk. The CC component then saves the metadata, securely erases
the keys from memory, and terminates execution. Finally, the SBC system
powers down.

\subsection{Key Management \label{keymanagement}}

Decryption and authentication keys can be supplied to the CC through various
mechanisms:

\begin{itemize}
	\item The keys may be stored directly within the CC’s internal
	storage if the user chooses to operate it similarly to a hardware
	authentication token (that is, the physical CC device itself acts
	as the key required to decrypt the external disk). In this case,
	the CC stores, as a configuration parameter, a key derivation
	function (KDF) output of a user-provided password specific to the
	disk. From this secret, sufficient keying material is securely
	derived for both encryption and authentication purposes.

	\item Similarly, this secret could be stored on an SD card or a
	USB flash drive, such that the physical medium itself is treated
	as the key.

	\item The host may provide the secret through an initial
	configuration protocol with the CC program. For instance, the
	first write operation performed by the host to a designated disk
	block (e.g., block 0) upon connection, containing a predefined
	magic number in its leading bytes, can be used to convey the
	secret. Note that this write operation is not a real disk
	operation, it does not affect the external disk. The written
	secret will not be stored in any storage, it will be used by
	the CC and removed from the main memory when the program finishes.

	This approach is analogous to the interface provided by some
	synthetic or virtual file systems, such as the \texttt{/proc}
	file system in Linux.

	 To use this method across any host operating system, it suffices
	 to implement a small program or script that retrieves the
	 password, executes the KDF and performs the initial write
	 operation to the block device. On Unix-like systems, this
	 operation can be achieved directly using the \texttt{dd} command.

\end{itemize}

Those mechanisms can be combined to provide multi-factor authentication.

\section{Implementation \label{impl}}

Our current prototype is a Go program of $\approx$ 3100 lines of code
and several shell scripts ($\approx$ 330 lines of code) for the service
and configuration.

To ensure maximal simplicity and adherence to established standards, the
proposed prototype adopts the AES encryption algorithm operating in CTR
mode. We could have used other modern approaches like authenticated
encryption (e.g. AES-GCM) or more efficient algorithms for implementations
relaying entirely on software, such as ChaCha20, but we chose to build
and evaluate an initial prototype that adheres as closely as possible to
popular encryption schemes.
%% V2
Note that changing the encryption and authentication algorithms
in our implementation is straightforward.

The same 256 bit key is used for all the blocks. Each block uses a
different random nonce\footnote{The CTR mode uses the term nonce instead
of \emph{initialization vector} (IV).}. Each time the block is written,
a new nonce is generated\footnote{Nonces are generated in the CC by
the CSPRNG provided by the Linux kernel.}. All the nonces are stored in
the CC metadata. The block size of the CC is configurable. By default,
it uses 4096 bytes (i.e. 256 AES blocks per disk block).

CTR is a malleable mode. Therefore, some kind of integrity or
authentication must be applied.  For block data authentication, the
prototype uses
%%V2
HMAC-SHA1\footnote{Note that HMAC-SHA1 is still considered secure, because
there is no indication that attacks on SHA-1 can be extended to it~\cite{hmacsha1}.}.
All blocks share the authentication key.
When a block is read, the HMAC of its cleartext is verified.  When a
block is written, the HMAC is updated.  The HMACs are stored together
with the nonces.

Since the nonce is updated on every block write operation, replay attacks
cannot be used to bypass authentication: if the attacker replaces the
ciphertext of a block with a previous version of that block, the HMAC will
not match, because the nonce used in the earlier version differs from the
current one. Consequently, the CC follows the FADE model and provides
the temporal data authentication property~\cite{BENADJILA2022100465}:
the user is always sure that she is working with the data stored by the
last previous legitimate write operation.

The size of the metadata depends on the disk and the block size: we need
to store a nonce (16 bytes) and a HMAC (20 bytes) per block. For instance,
for a 480GB external disk with 4096 byte blocks, we need $\approx$ 4GB
of metadata.  Nowadays, this amount of storage is relatively cheap. A
12\euro{} 64 GB MicroSD card would provide the metadata for 16 disks
like this one.

Our prototype has not undergone fine-grained optimization. It processes
the NBD petitions sequentially.  The metadata (nonces and HMACs)
is cached in memory, in a hash table indexed by the block address
(specifically, a Go map).  The modified metadata is constantly synced
to storage by a dedicated thread.  In addition, all the memory of the
process is \emph{mlocked}, that is, the memory pages of the process
can't be evicted from main memory (to the swap partition).

%%V2
\subsection{Data Corruption (Crashes and Attacks)}

Ciphertext and its corresponding metadata (nonce and HMAC) are stored
in two different devices (the untrusted disk and the CC), therefore
their updates are not atomic.
The metadata becomes corrupted, for example, by a system
crash that interrupts while a write operation is being performed.
This situation can leave the encrypted data on disk
inconsistent with the CC metadata. In this case,
the affected blocks can no longer be authenticated, provoking I/O errors,
as stated in Section ~\ref{threatmitigation}.

To recover from such a failure, we provide a tool
that permits the user to manually reset the state for incoherent blocks
after a crash.
Data being written at the time of the crash is lost because it
did not reach the disk. The host
file system is responsible for handling such failures as it would with
a conventional hard drive, using mechanisms such as file system journals
or soft updates. The recovery tool ensures that subsequent accesses do not result
in I/O errors due to the crash.

In the case of a disk rollback attack (where the attacker replaces disk
blocks with previous versions of those same blocks), replay attacks,
or any other kind of substitution of the encrypted data,
the situation is indistinguishable from the previous case (the inconsistency
is caused by a crash): the encrypted data in these blocks will not correctly
authenticated with the current metadata stored in the CC. Thus, the tampered
blocks will be treated as invalid blocks, resulting
in I/O errors. The user will detect the disk malfunction and can use the
aforementioned recovery tool, if necessary, to reset the affected blocks
and eliminate the errors.

\subsection{Address Permutation}

As explained in the threat model, in order to mitigate the risk of
cryptanalysis of a disk block that may contain known plaintext (e.g. the
blocks where the primary boot loader may be stored), the CC performs a
keyed block address translation: the block requested by the host does
not correspond to the same address as the block on the hard disk that
contains the encrypted data.  Without the key, the attacker is not able
to locate those sensitive blocks in the encrypted disk.

This can be implemented in different ways. Currently, our prototype
includes two methods to perform the keyed permutation: \footnote{Note
that the permutation has to be calculated independently for each index
(what is sometimes called an oblivious permutation), i.e. without keeping
any other extra state around, like the list of all the permuted indexes
(which would occupy too much space).}:

\begin{enumerate}
	\item Custom method. This method is simple and fast.  Given the
	size (i.e. the addresses space) $s$, it is divided in two
	overlapping parts by choosing a prime $p$, with $s >= p >
	s/2$. Then a random permutation $P_r(a,b)$ is picked by choosing
	randomly $a$ and $b$ with $p > a > 1$ , $p > b >= 1$ and computing
	$(aI+b)\%n$, where $I$ is the index.  The existence of an inverse
	for the group guarantees that $P_r(a,b)$ is a permutation. The
	final permutation $P$ is a combination of $N$ rounds. Each round
	is $P_r$ followed by a Faro shuffle (which is non-linear and
	makes inverting $P$ more difficult). The values of $(p, a, b)$
	are secret (i.e. they are the permutation key).

	\item Thorp shuffle method~\cite{10.1007/s00145-017-9262-z}, where the random
	bits which control the shuffle are calculated deterministically from
	the key, the index
	and the round of shuffle using an HMAC like so:
	$HMAC(key, index | round)$.  This method is more secure and costly: it is
	40 times slower to calculate than the custom permutation but
	provides strong security guarantees.
\end{enumerate}

In general, the permutation impacts system performance, as will be shown
in the evaluation. This is related to sequential disk access, and space
locality in the disk access for which both the disk firmware and the CC
operating system are highly optimized for. Strategies like read-ahead
and caching cease to be effective on both sides of the CC-disk interface.

\subsection{Ghost Reads}

A malicious disk firmware could easily defeat the aforementioned
mitigation, since certain disk blocks are accessed following a
well-defined temporal pattern. In this way, the malicious firmware can
already determine which physical disk block corresponds to address 0 as
seen by the host. Such firmware could then maintain a log of read and
write operations to infer the addresses of other sensitive disk blocks
(e.g., the FAT table, inode vectors, file system journal, etc.).

%%V2
To mitigate this risk for read-only relevant blocks,
the CC can be configured to perform what we refer
to as \emph{ghost reads} over certain regions of the address space used
by the host. When the host needs to read a block within one of these
designated regions, the CC actually performs the read operation for
the requested block along with a set of fake reads that are discarded
transparently to the host. For a given block, the same number of dummy
reads is always performed, targeting the same set of blocks and in the
same order. The number of dummy reads is configurable ($NG$).

%%V2
Note that this mitigation should be used only on read-only
blocks\footnote{If the block is writable, this mitigation provides no benefit.},
when the address mapping is clearly broken by the operating system
or application behavior. For
example, if the operating system always reads the block 0 when a disk
is attached (this is an usual behavior in Linux and the cleartext of
this block may be partially known by the attacker), this mitigation may
be effective: with it, the attacker has to analyze $NG$ blocks; without
it, the attacker already knows the address of block 0. In other cases,
this mitigation should be applied with considerable caution.

To achieve this, a dedicated thread (i.e. a Go \emph{goroutine})
is spawned to execute a function that is carefully designed to remain
subtle to avoid timing attacks:

\begin{enumerate}
	\item Generate a fixed pseudo-random list of \emph{ghost blocks} for
	a block. The generator
	uses as seed the sum of two numbers: the address of the
	corresponding block and a long secret number for this
	disk (which is part of the configuration parameters).

	\item Select a random position of the target block within
	this sequence.	The position is also extracted from the same
	pseudo-random generator.

	\item Perform all the reads sequentially, forwarding only the
	result of the correct block to the main thread (via a Go channel)
	and discarding the rest. Thus, from the correct read to the last
	ghost read of the sequence, the read operations are asynchronous
	(that is, the main thread does not wait for them to finish).

\end{enumerate}

Obviously, this approach also entails a substantial performance overhead,
as it significantly increases the number of random read operations. As
with the previously mitigation, it can be selectively disabled.

%%V2
Note that the ordered list of ghost blocks is constant for a block. Thus,
the attacker has no information to recover
sensitive block locations after observing a certain number of
accesses for the same blocks,
except when one of the \emph{ghost blocks} is already used by the
host file system (i.e. there is a \emph{real}
operation over a block that was randomly selected as a \emph{ghost block}).
In this case, the total count of reads for this \emph{ghost block}
would be greater than
the count for the other.
This information permits the attacker to discard this very \emph{ghost block}
as the relevant one (the other \emph{ghost blocks} are still valid).

To avoid this leakage,
the CC could generate additional ghost reads for the rest of \emph{ghost blocks}
of the corresponding list.
This way, the CC could equalize the access counts among the blocks designated as ghost
reads. However, this could impose excessive overhead on the system.

%%V2
In the current implementation, the user can set $NG$, considering the probability
of losing a number of \emph{ghost blocks} as a result of this issue.
The probability of selecting a \emph{ghost block} already used by the file system
as a \emph{ghost block} for a region depends on the disk usage ratio and the
size of the disk.
Being $U$ the usage ratio of the disk, $T$ the number of blocks of the disk,
and $R$ the total amount of blocks with \emph{ghost reads} (i.e.
the sum of the regions), the probability can be estimated
as $(NG * R) / (T * U)$\footnote{Note that $NG*R \leq T*U$ as we consider
the ghost blocks in use, so this is indeed a probability.}.
For example, a 480 GB device with a disk usage of 75\%, 17 regions of 32 KiB
for ghost reads, and $NG=100$ (the same configuration evaluated in
Section \ref{eval}), the probability is $\approx 0.0001$. Note also that
the same block could be selected as \emph{ghost block} twice, but the probability
is very low ($\approx 8.5e-9$ for this example).

%% >>> T=480 * 1000 * 1000 * 1000 / 4096
%% >>> R=17*32*1024/4096
%% >>> NG=100
%% >>> U=0.75
%% >>> (NG * R) / (T*U)
%% 0.0001547377777777778

%%V2
In any case, ghost reads should be regarded only as an additional mitigation,
not a secure defense, for our system.
Note that, with standard FDE schemes, an attacker controlling the disk
firmware always knows exactly which blocks are the sensitive ones.
Moreover, in the case of self-encrypted disks (SED), the disk firmware must be
fully trusted because it handles clear data and keys.

\section{Evaluation \label{eval}}

All the experiments have been executed on the following hardware:

\begin{itemize}
	\item A host machine with a $12^{th}$ Intel Core i7-1280P processor, 20 cores
	and 48 GiB of RAM, running Ubuntu Linux 24.04 (6.14.0-37-generic kernel).

	\item A CC running on an Orange Pi 5 Ultra SBC (64-bit mode) with a
	quad Cortex-A76 at 2.2 2.4GHz and a quad Cortex-A55 at 1.8GHz, and 16
	GiB of RAM.
	This SBC has a USB 3.0 OTG port (super-speed).
	The operating system is Armbian Linux 25.5.2 (6.1.115 kernel).

	\item As the reference external drive, we selected a 480 GB
	Kingston A400 SSD mounted in an 2.5 inch SATA-to-USB 3.0/3.1
	Gen 1 external enclosure (model TQE-2527B).
\end{itemize}

\begin{figure}[ht]
\begin{center}
	\resizebox{0.8\columnwidth}{!}{
		\includegraphics{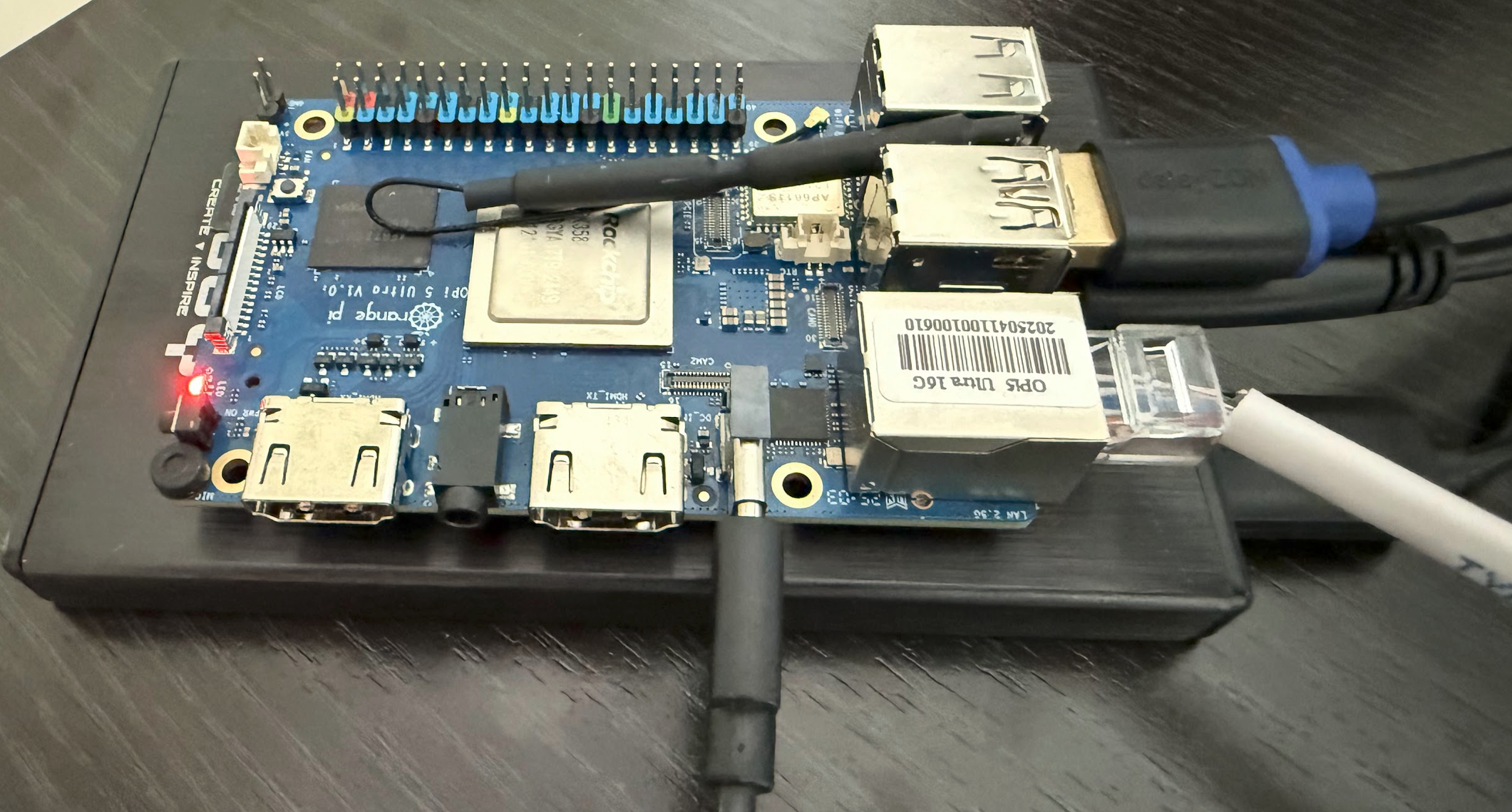}
	}
	\caption{The Orange Pi board over the external USB drive
	used in the experiments. \label{orange}}
\end{center}
\end{figure}

Figure~\ref{orange} shows the SBC and the drive.
We conducted several experiments using two different standard I/O
benchmarks to evaluate the approach and our first prototype:
FIO~\cite{fio} and Filebench~\cite{tarasov2016filebench}.

\subsection{FIO}

The goal of the FIO experiments is to evaluate the performance of raw
I/O of our approach, comparing it with other setups. The experiments
were conducted under different scenarios:

\begin{enumerate}
	\item The external disk directly connected to the host
	(label \texttt{device-kingstonA400}).
	\item The external disk formatted as a encrypted LUKS volume
	(label \texttt{luks-kingstonA400}).
	In this case, we measure the access to \texttt{/dev/md-0}, the mapped device
	that represents the cleartext contents of a LUKS volume.
	\item The external disk connected to the SMB and directly exported to the host
	via OTG (label \texttt{device-otg}). In this case, the SBC is just a relay
	computer exporting the disk (i.e. \texttt{/dev/sda} in the Orange Pi)
	through OTG to the host.
	That is, in this scenario the CC software is not running.
	This scenario lets us measure the cost of the OTG mechanisms.
	\item The CC encrypting/decrypting the external disk,
	configured without our custom permutation (label \texttt{cc-noperm}).
	\item The CC encrypting/decrypting  the external disk,
	configured with our custom permutation (label \texttt{cc-perm}).
\end{enumerate}

In all cases, we consider a block size of 4096 bytes. We have measured the
bandwidth and the latency
for all these scenarios.

Since the cost of ghost reads is fixed (i.e. multiplying the number of
read operations by $NG$
in disk areas where this mitigation is enabled), this feature
was excluded from the FIO evaluation.

\subsubsection{Bypassing the host OS cache}

\begin{figure}[ht]
\begin{center}

	\resizebox{0.4\columnwidth}{!}{
		\includegraphics{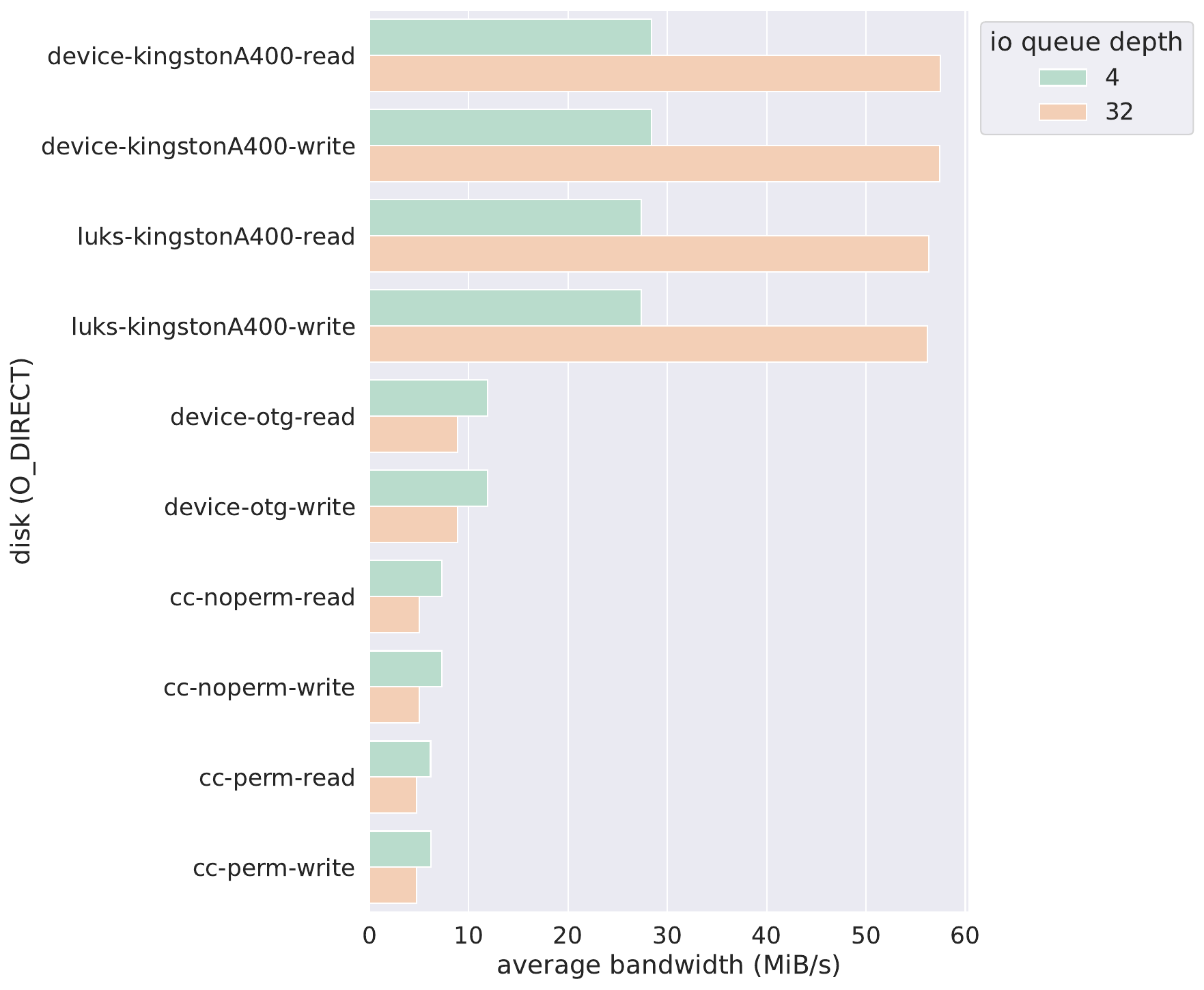}
	}
	\resizebox{0.4\columnwidth}{!}{
		\includegraphics{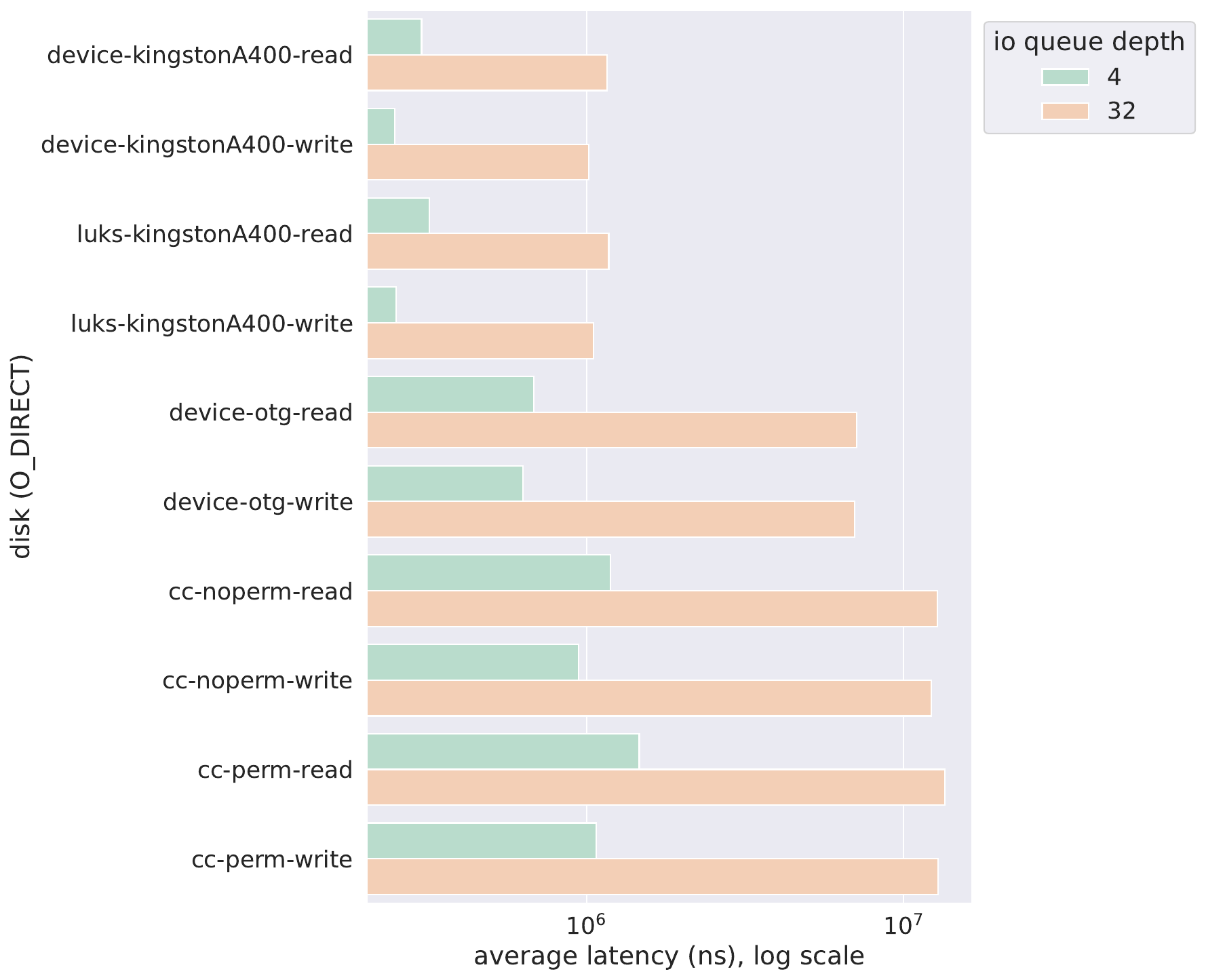}
	}
	\caption{Bandwidth and latency for mixed random read and
	write operations (host OS cache disabled). \label{odirect-mixed}}
\end{center}
\end{figure}

\begin{figure}[ht]
\begin{center}
	\resizebox{0.4\columnwidth}{!}{
		\includegraphics{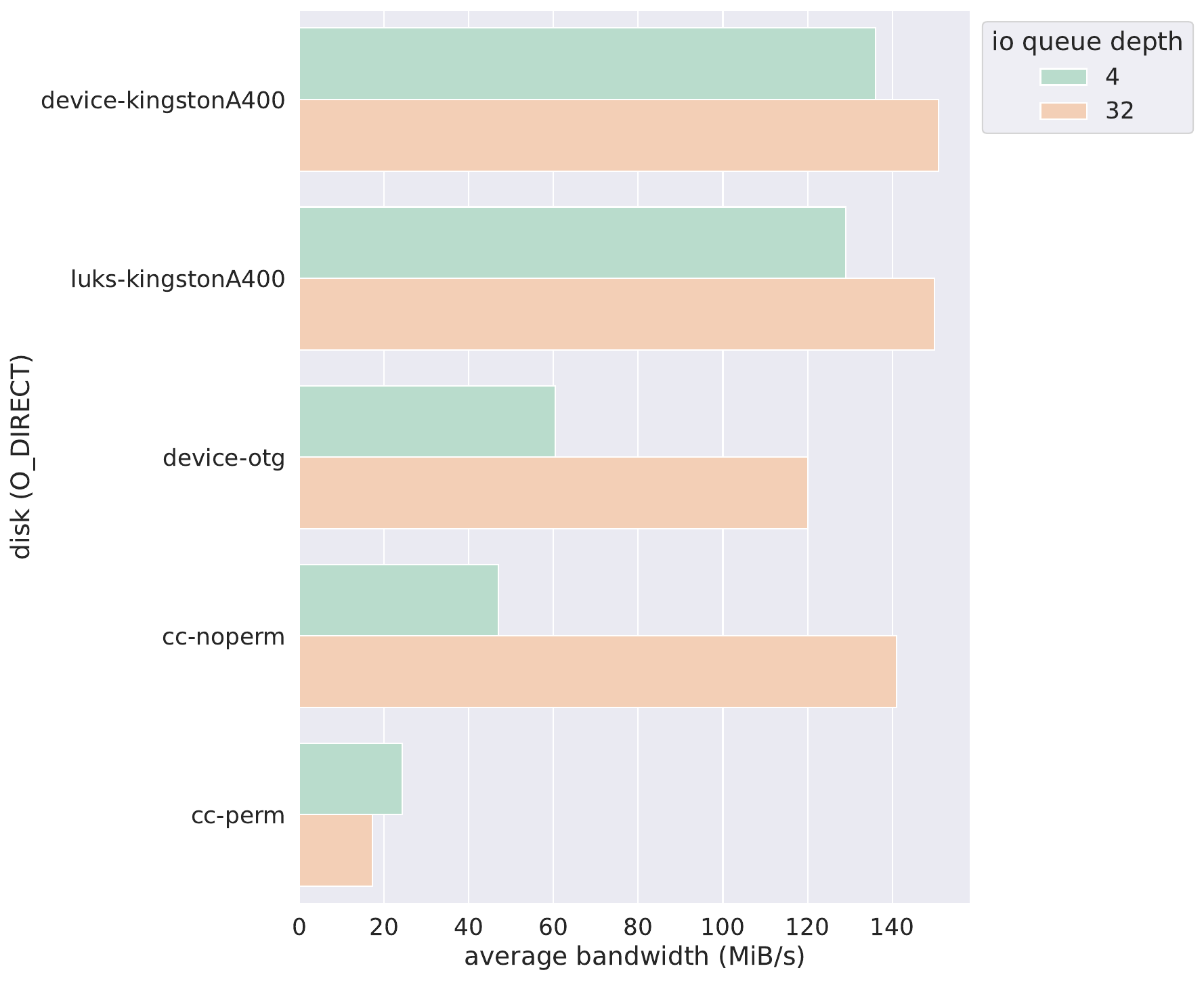}
	}
	\resizebox{0.4\columnwidth}{!}{
		\includegraphics{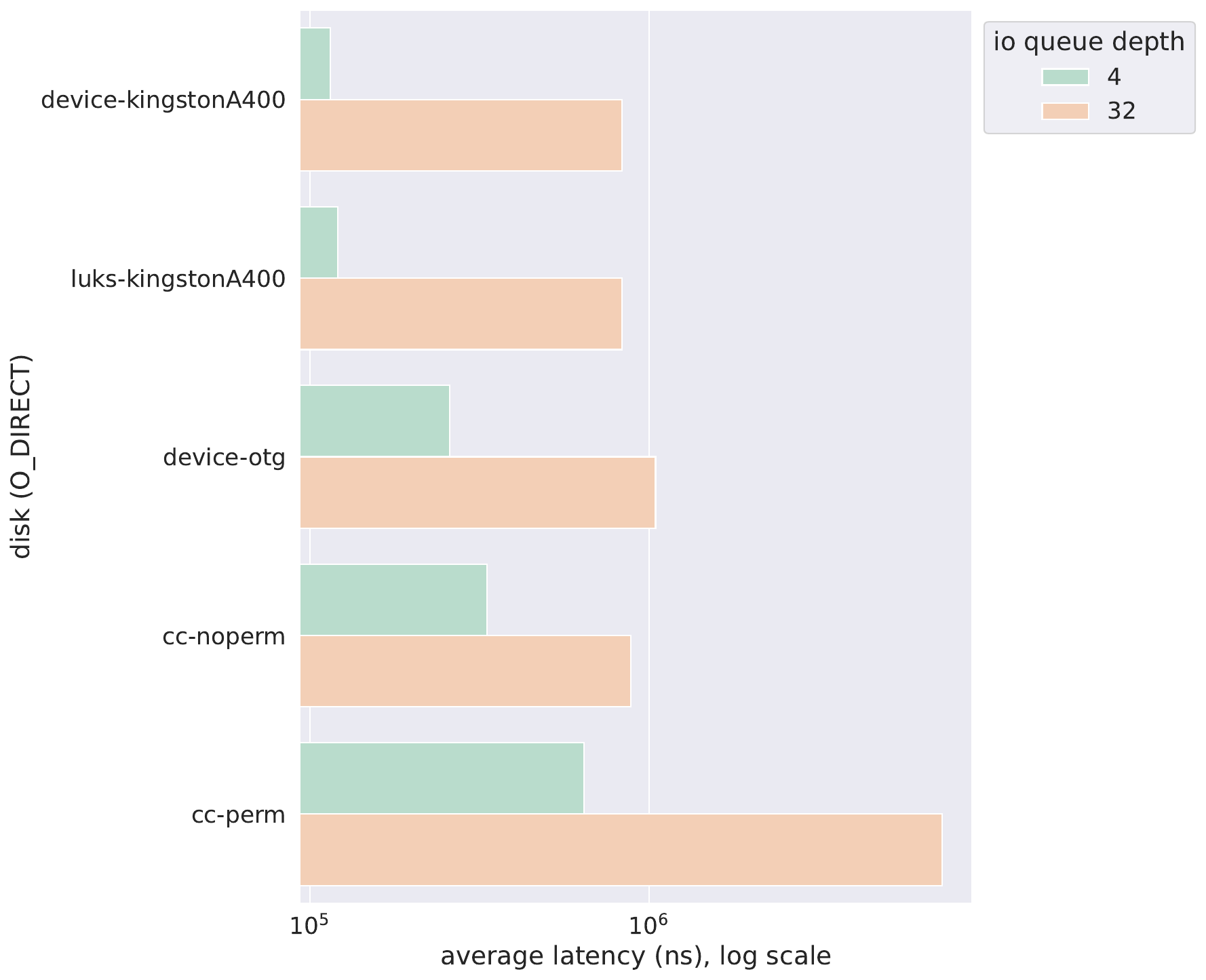}
	}
	\caption{Bandwidth and latency for sequential read operations
	(host OS cache disabled). \label{odirect-seqread}}
\end{center}
\end{figure}

\begin{figure}[ht]
\begin{center}

	\resizebox{0.4\columnwidth}{!}{
		\includegraphics{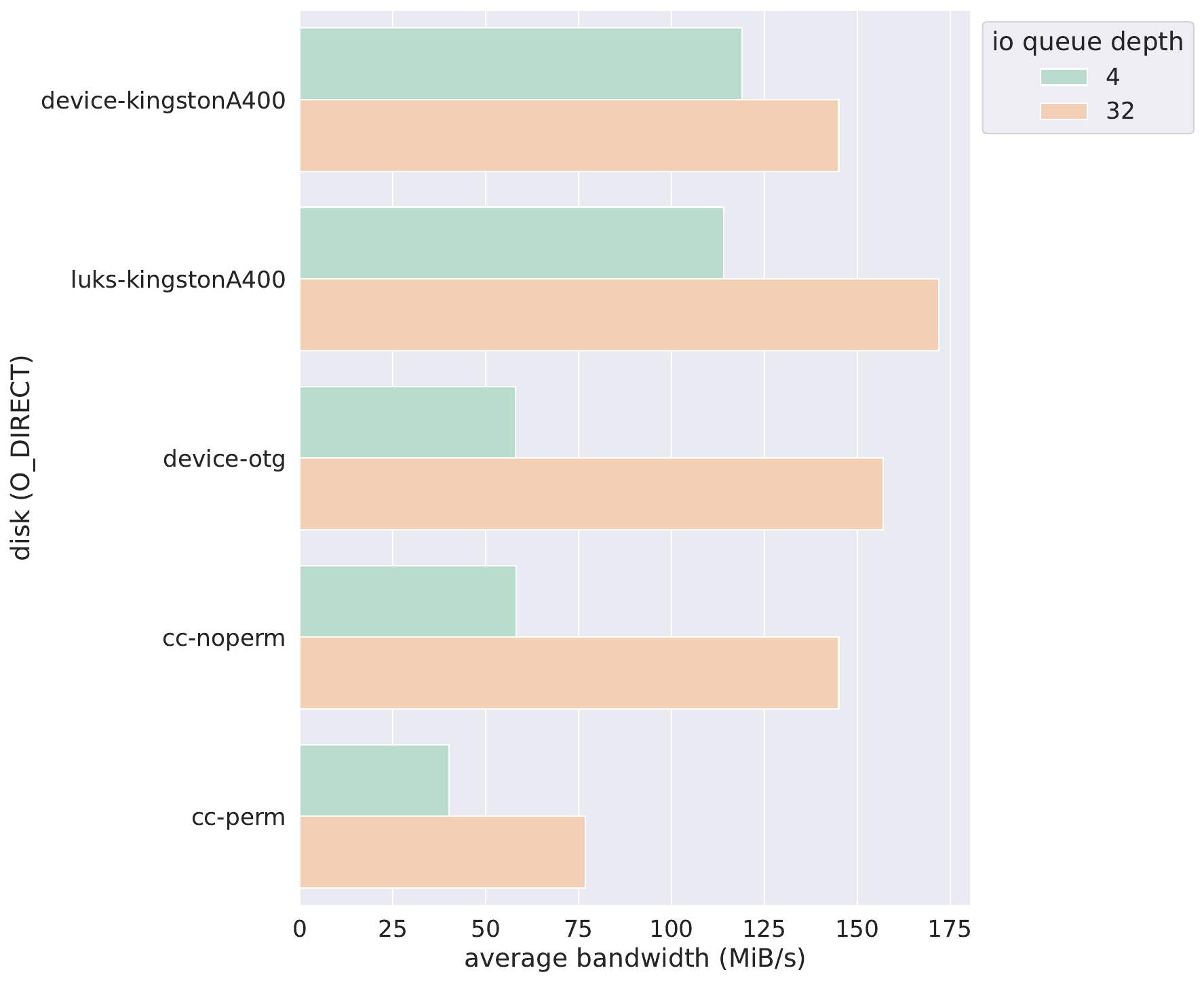}
	}
	\resizebox{0.4\columnwidth}{!}{
		\includegraphics{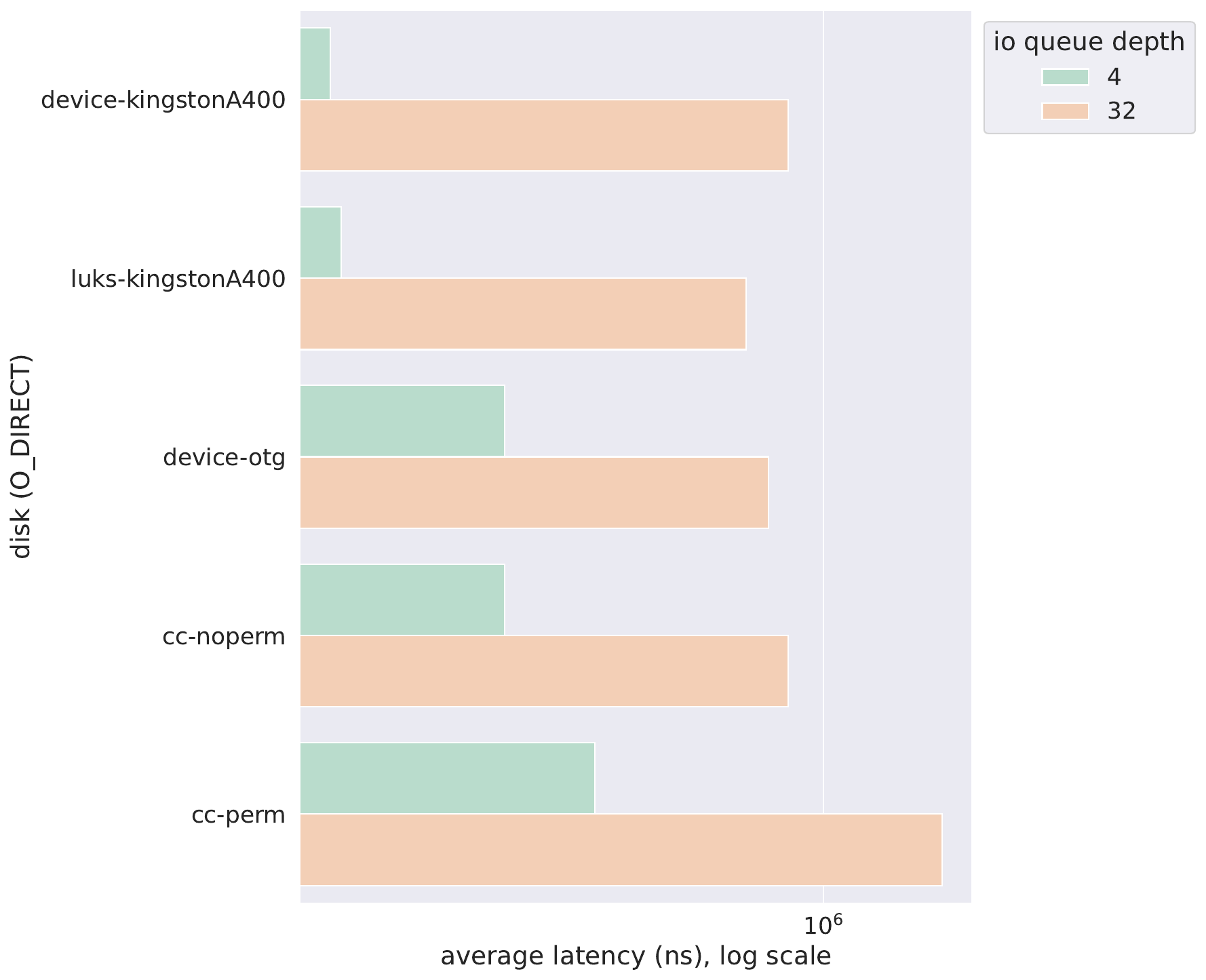}
	}
	\caption{Bandwidth and latency for sequential write operations
	(host OS cache disabled). \label{odirect-seqwrite}}
\end{center}
\end{figure}

\begin{figure}[ht]
\begin{center}
	\resizebox{0.4\columnwidth}{!}{
		\includegraphics{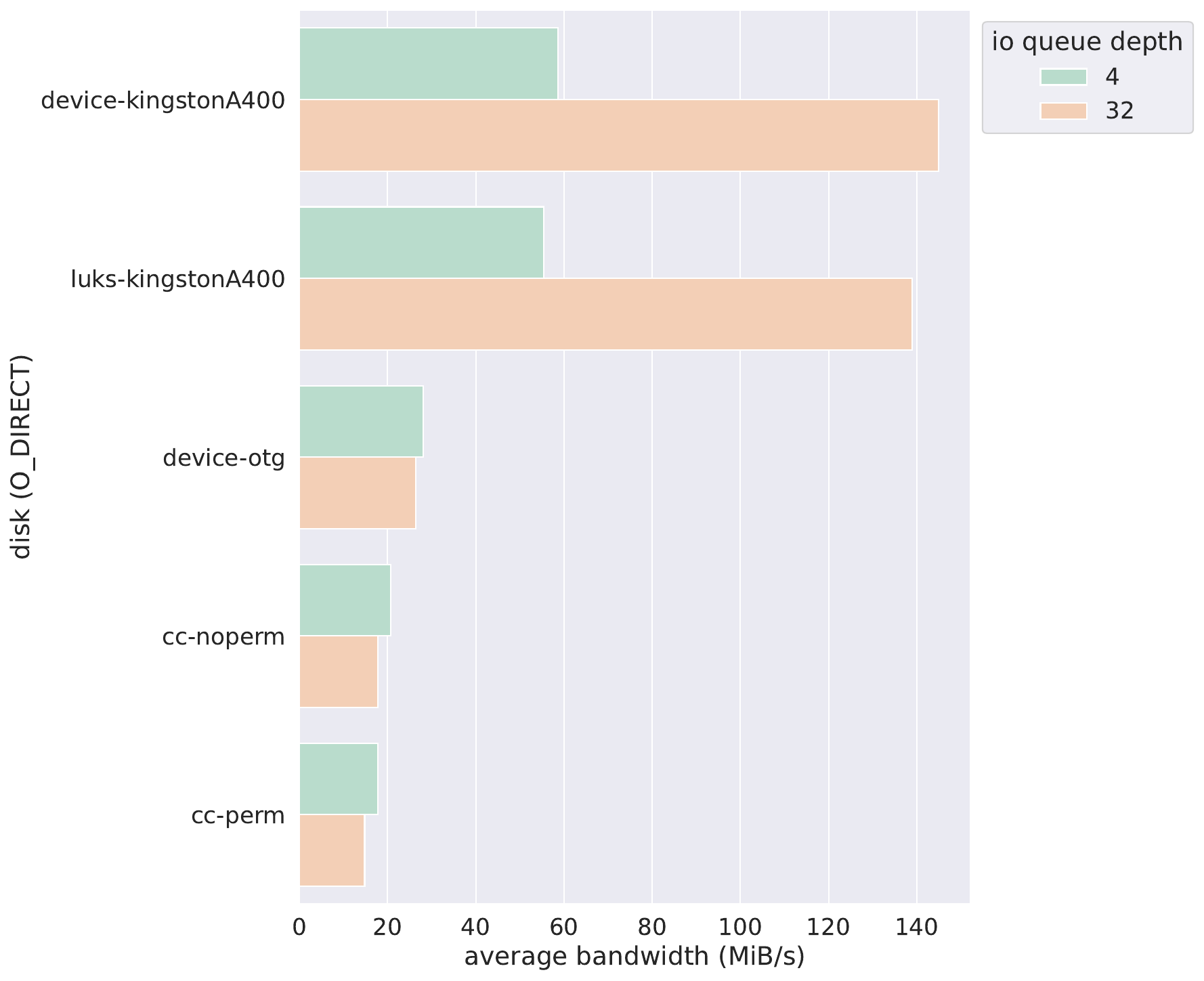}
	}
	\resizebox{0.4\columnwidth}{!}{
		\includegraphics{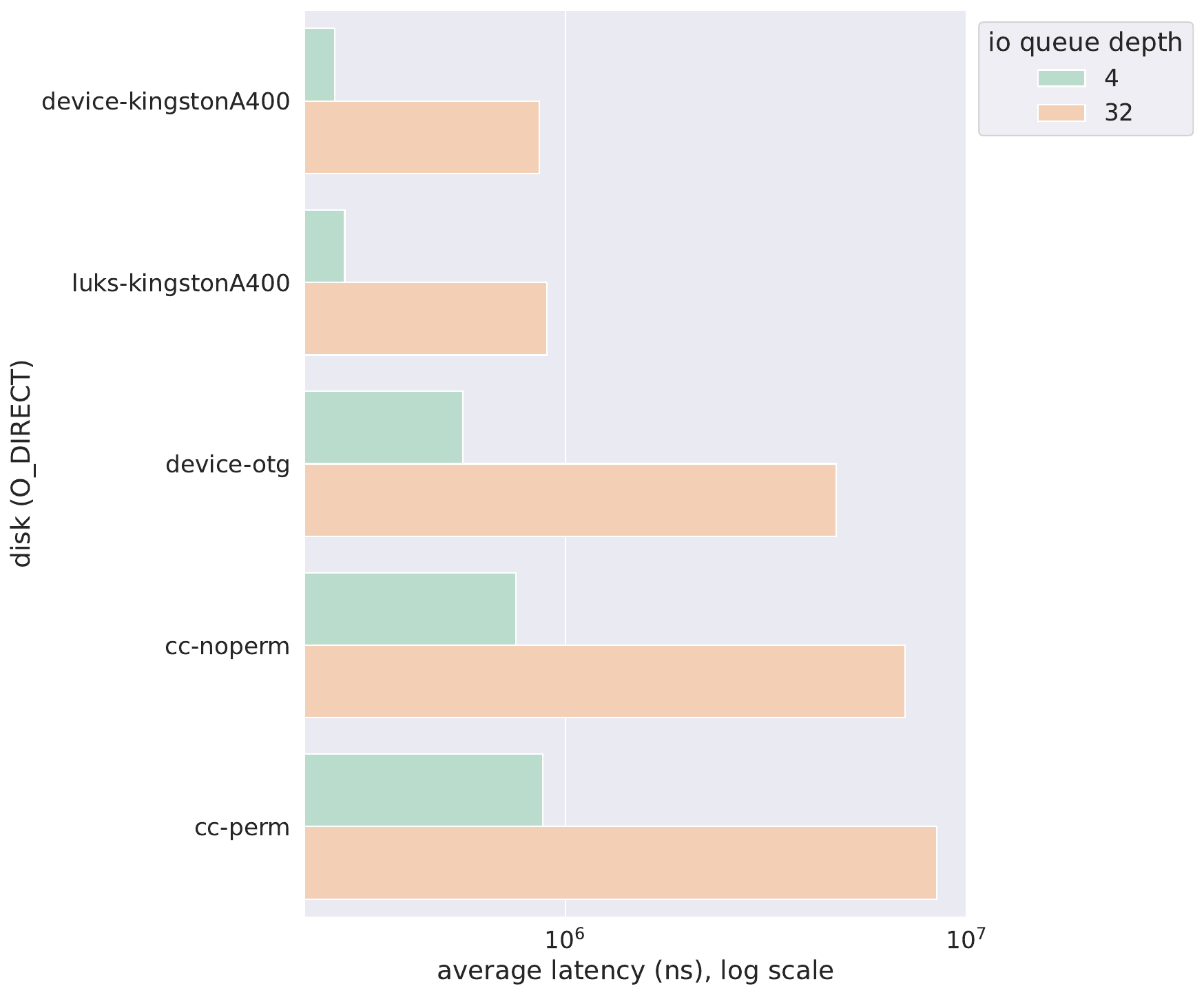}
	}
	\caption{Bandwidth and latency for random read operations
	(host OS cache disabled). \label{odirect-ranread}}
\end{center}
\end{figure}

\begin{figure}[ht]
\begin{center}

	\resizebox{0.4\columnwidth}{!}{
		\includegraphics{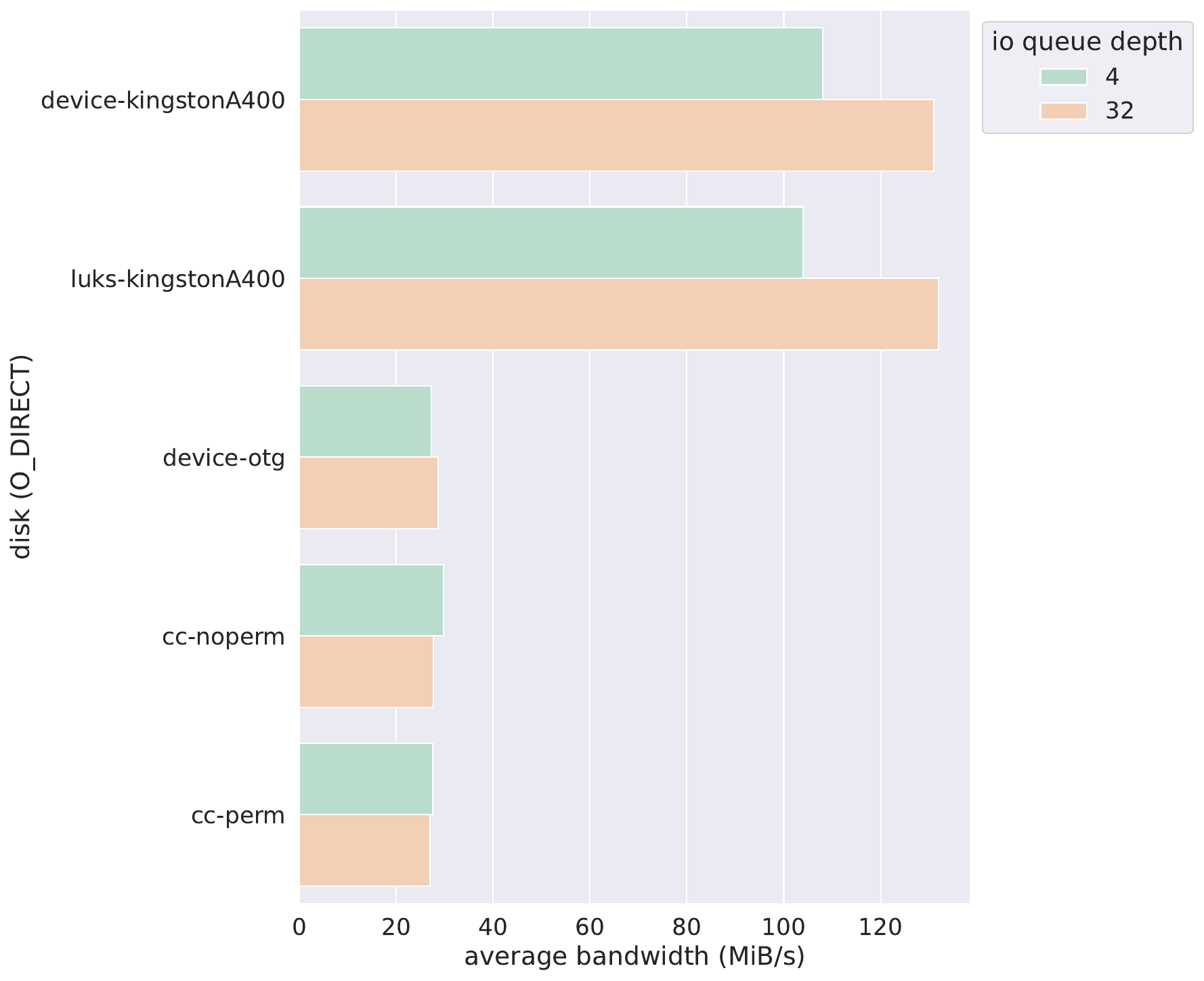}
	}
	\resizebox{0.4\columnwidth}{!}{
		\includegraphics{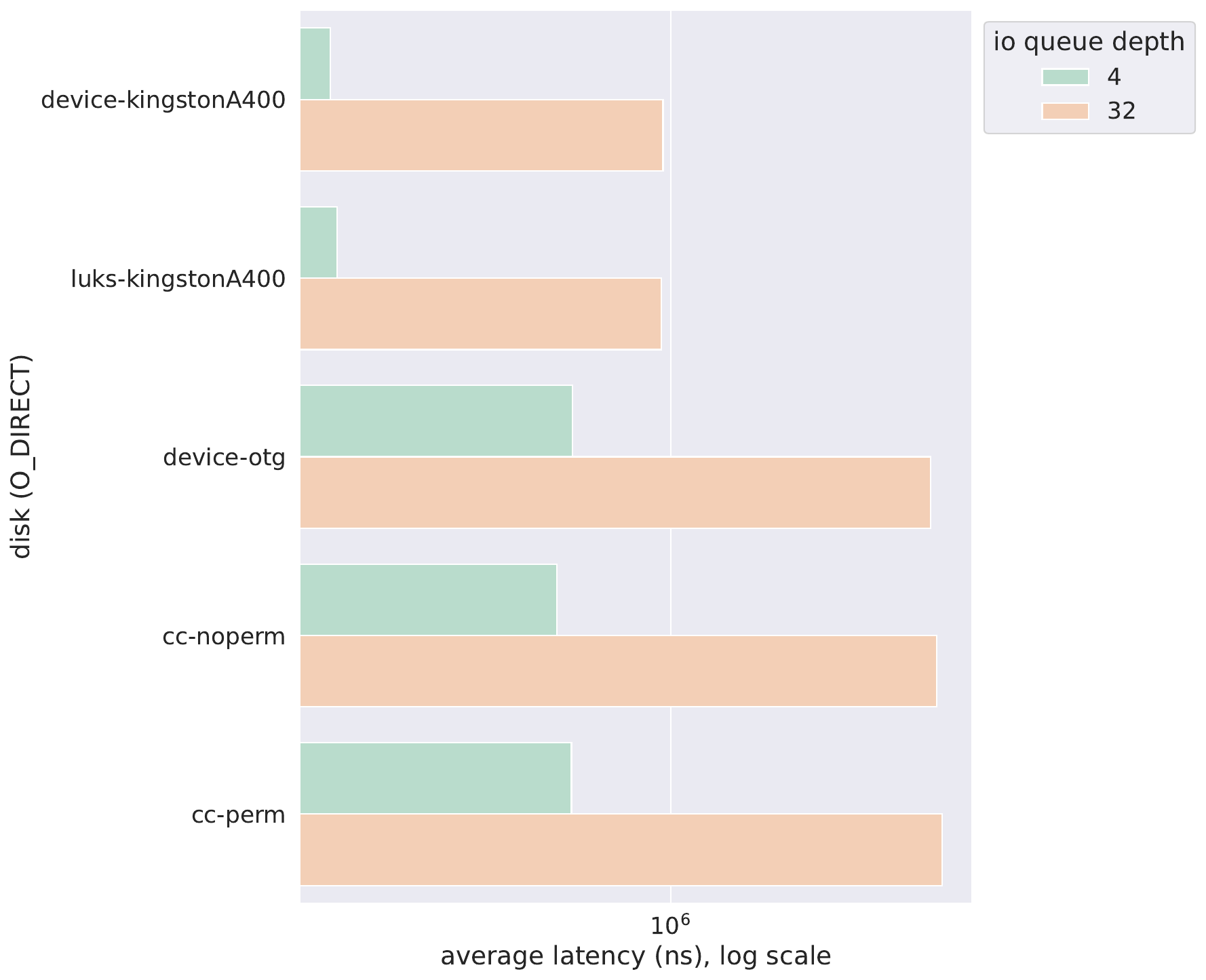}
	}
	\caption{Bandwidth and latency for random write operations
	(host OS cache disabled). \label{odirect-ranwrite}}
\end{center}
\end{figure}

To ensure accurate measurement of device performance, FIO was
configured to disable the host operating system’s cache. That is,
read and write operations are performed directly on the device using the
\texttt{O\_DIRECT} flag of the \texttt{open} system call.  For these
experiments, asynchronous I/O was evaluated (using \texttt{libaio}).
We configured two different I/O queue depths: 4 and 32.  This parameter
specifies the number of concurrent in-flight operations that can be
issued simultaneously.  The experiments are time based (60 seconds,
with an extra warming time of 10 seconds).
Figure~\ref{odirect-mixed} shows the results for mixed random read
and write operations (50/50).
Figure~\ref{odirect-seqread} shows the results for sequential read operations.
Figure~\ref{odirect-seqwrite} shows the results for sequential write operations.
Figure~\ref{odirect-ranread} shows the results for random read operations.
Figure~\ref{odirect-ranwrite} shows the results for random write operations.

\subsubsection{Using the host OS cache}

\begin{figure}[ht]
\begin{center}

	\resizebox{0.4\columnwidth}{!}{
		\includegraphics{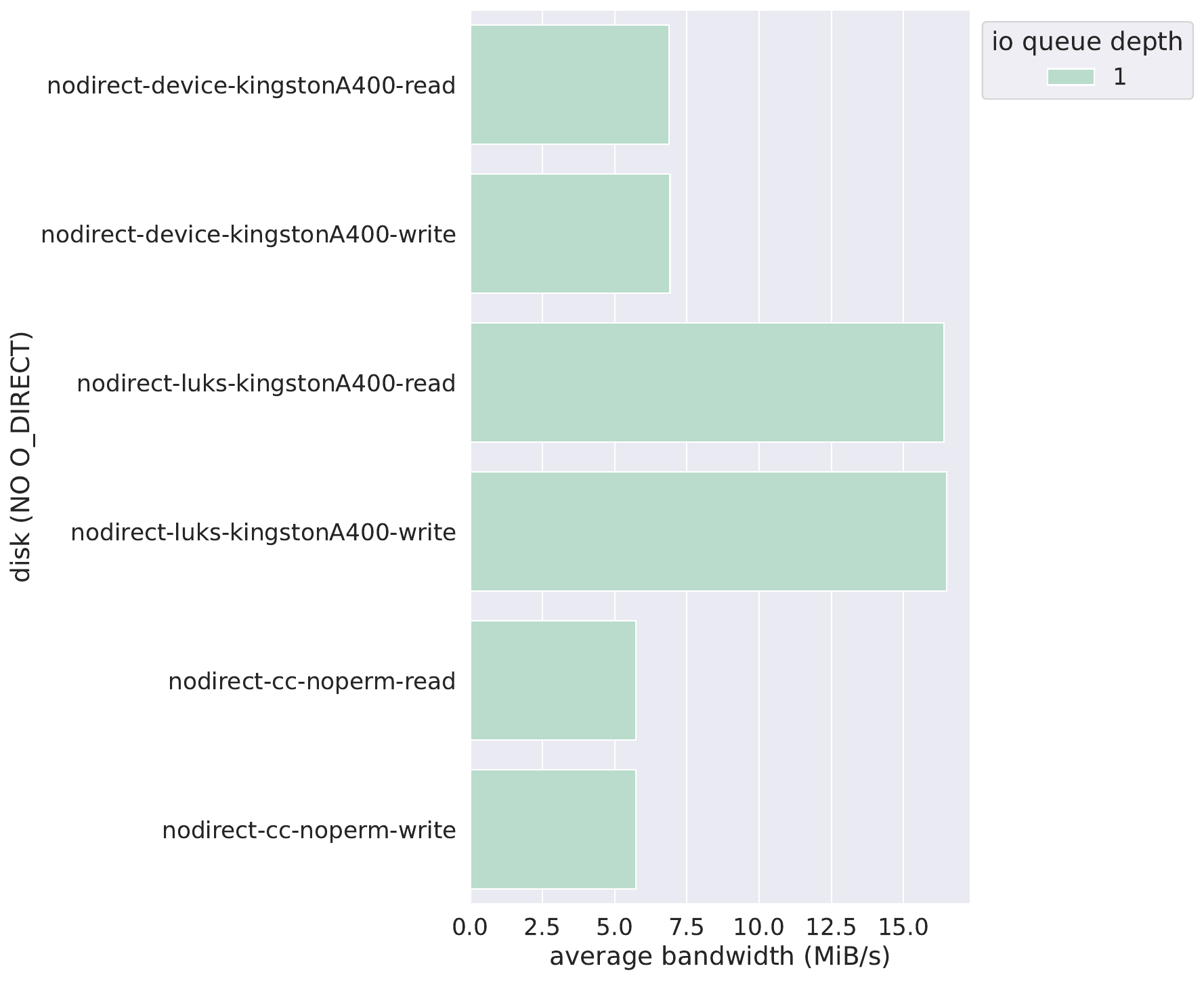}
	}
	\resizebox{0.4\columnwidth}{!}{
		\includegraphics{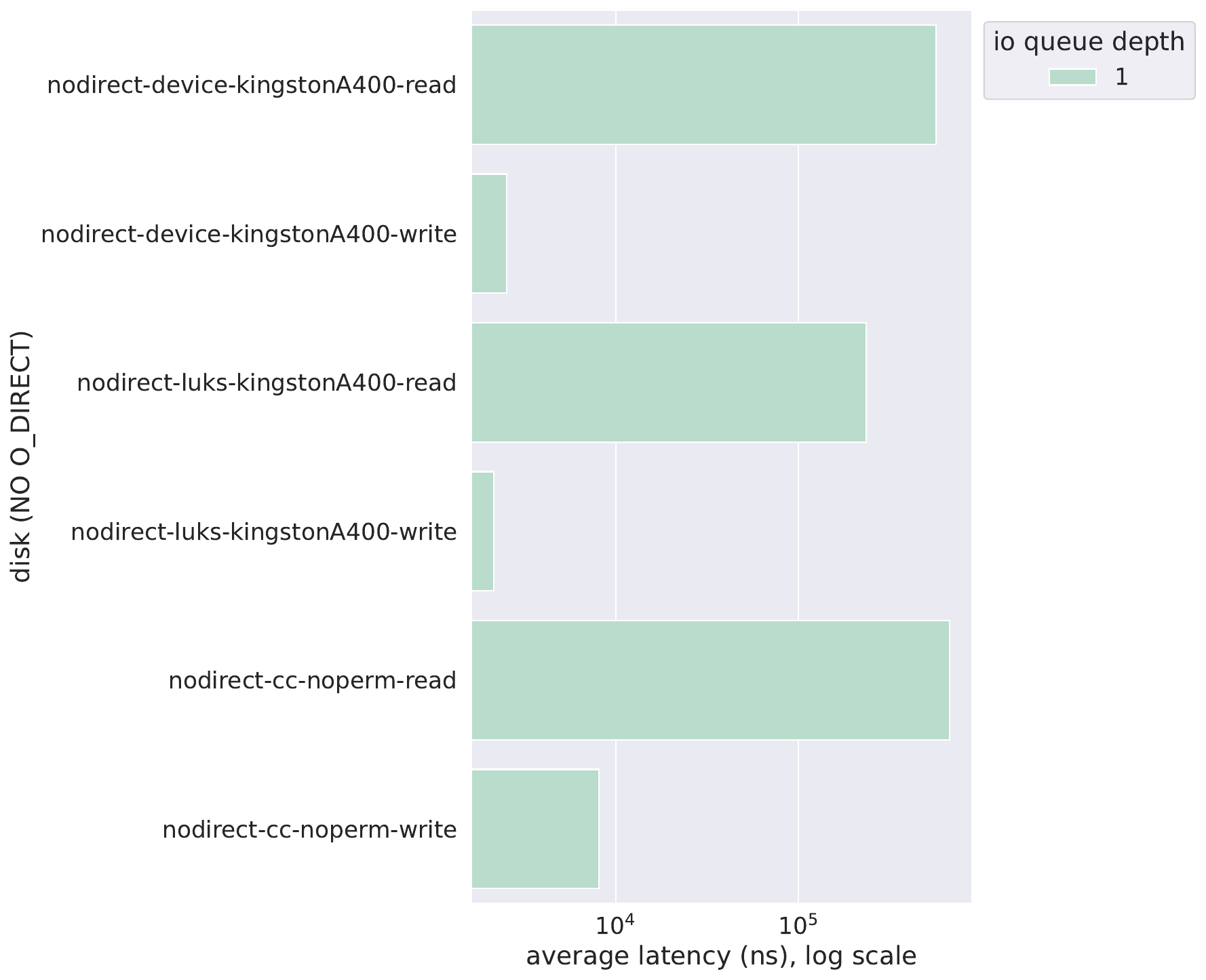}
	}
	\caption{Bandwidth and latency for mixed random read and
	write operations (host OS cache enabled). \label{nodirect-mixed}}
\end{center}
\end{figure}

\begin{figure}[ht]
\begin{center}
	\resizebox{0.4\columnwidth}{!}{
		\includegraphics{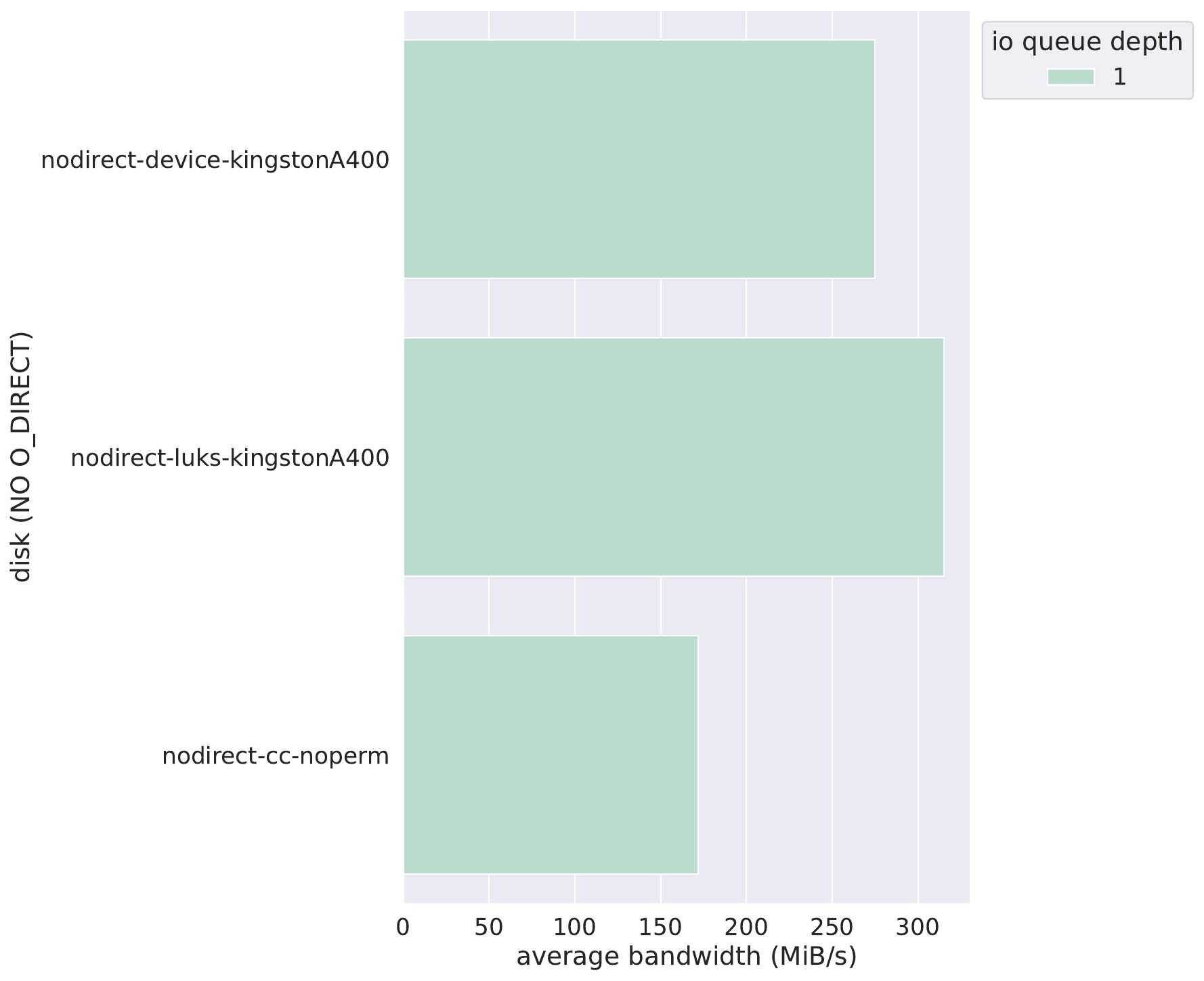}
	}
	\resizebox{0.4\columnwidth}{!}{
		\includegraphics{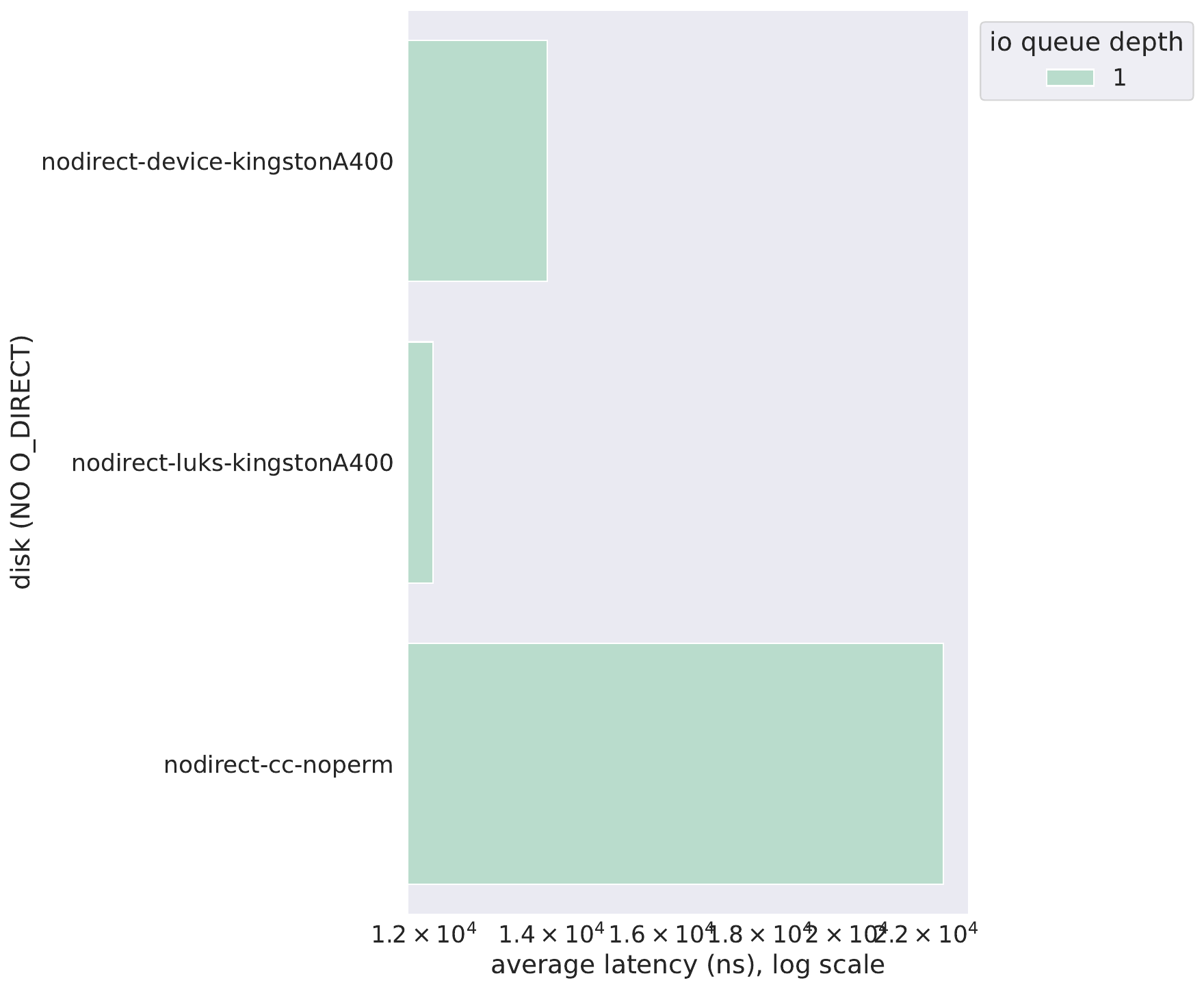}
	}
	\caption{Bandwidth and latency for sequential read
	operations (host OS cache enabled). \label{nodirect-seqread}}
\end{center}
\end{figure}

\begin{figure}[ht]
\begin{center}

	\resizebox{0.4\columnwidth}{!}{
		\includegraphics{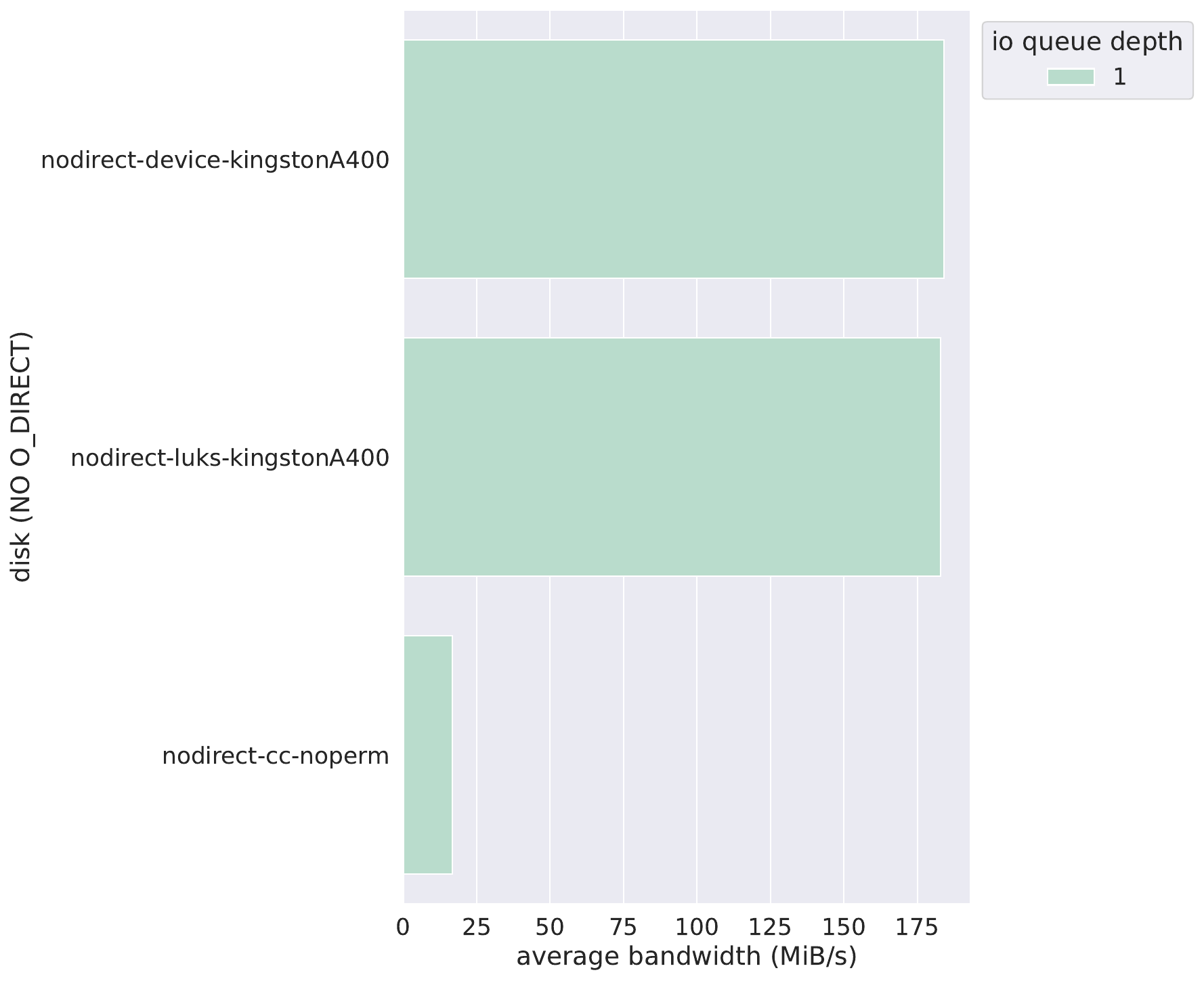}
	}
	\resizebox{0.4\columnwidth}{!}{
		\includegraphics{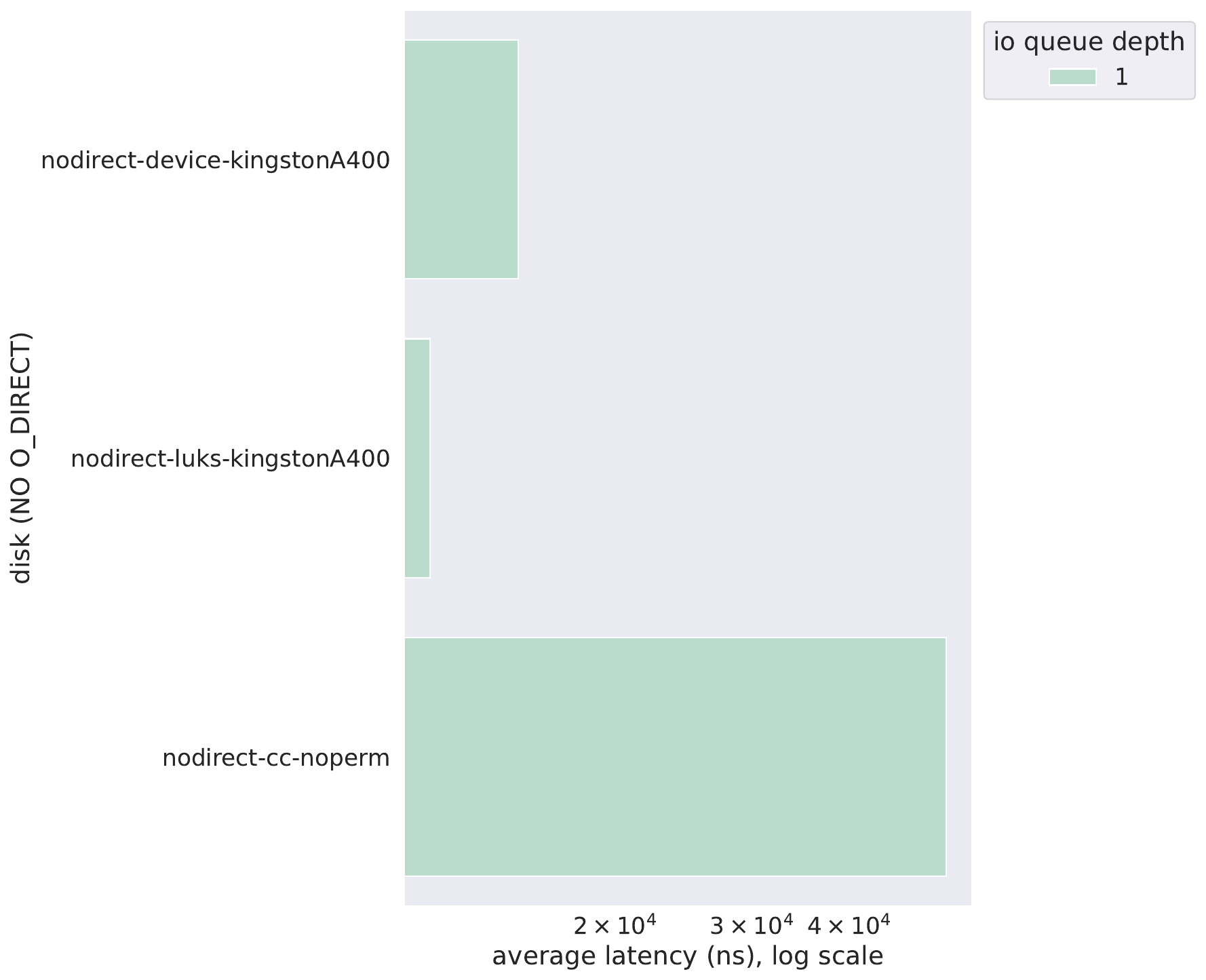}
	}
	\caption{Bandwidth and latency for sequential write
	operations (host OS cache enabled). \label{nodirect-seqwrite}}
\end{center}
\end{figure}

We have also performed some experiments with the host OS enabled, that is,
without the \texttt{O\_DIRECT} flag of \texttt{open}.  We have measured
three scenarios the disk directly connected to the host, LUKS, and the
most favorable scenario for our system (i.e. without permutation).

Note that, in this case, we are not measuring just the external device:
We are measuring the interaction of the device and the Linux kernel
cache mechanisms for disk blocks.  The memory capacity of the host (48
GiB) is sufficient to prevent cache pressure under the experimental
conditions considered.  In this case, the I/O queue size is 1.
Figure~\ref{nodirect-mixed} shows the results for mixed random read
and write operations (50/50).  Figure~\ref{nodirect-seqread} shows the
results for sequential read operations.  Figure~\ref{nodirect-seqwrite}
shows the results for sequential write operations.

\subsection{Filebench}

These experiments aim to measure the performance of a
conventional file system on the disk, unlike the previous
experiments conducted with FIO.

We conducted measurements using Filebench with two different representative
workloads across the following scenarios: direct disk access,
LUKS, CC without permutation, CC with permutation, and CC with permutation
and ghost reads. Note that, for these experiments, quantifying the overhead
of ghost reads
is both relevant and appropriate, particularly in regions where multiple
read operations are expected to occur during normal file system usage.

In all cases, the disk has been formatted with an Ext4 file system with
the default parameters in an Ubuntu Linux: 117212886 4KiB blocks,
29310976 inodes and journaling. For the ghost reads scenario, $NR$ is $100$ and
configured for 17 disk areas of 32 KiB
(the first blocks of the disk and all the copies of the superblock).
Before each experiment, the CC was restarted and the file system was formatted.

As stated before, we used two standard Filebench workloads:

\begin{itemize}
	\item \texttt{fileserver}, which is recommended for evaluating general
	file system performance.
	This workload performs different operations (\texttt{stat},
	\texttt{open}, \texttt{close},
	and \texttt{create}). It creates 1000 files and 50 threads,
	and uses 1 MiB I/O buffers.

	\item \texttt{webserver}, which is recommended for
	measuring read-heavy workload.  It also
	writes a log (\texttt{append} operation). In this case, the workload
	creates 100 threads and 1000 files.
\end{itemize}

\begin{figure}[ht]
\begin{center}

	\resizebox{0.7\columnwidth}{!}{
		\includegraphics{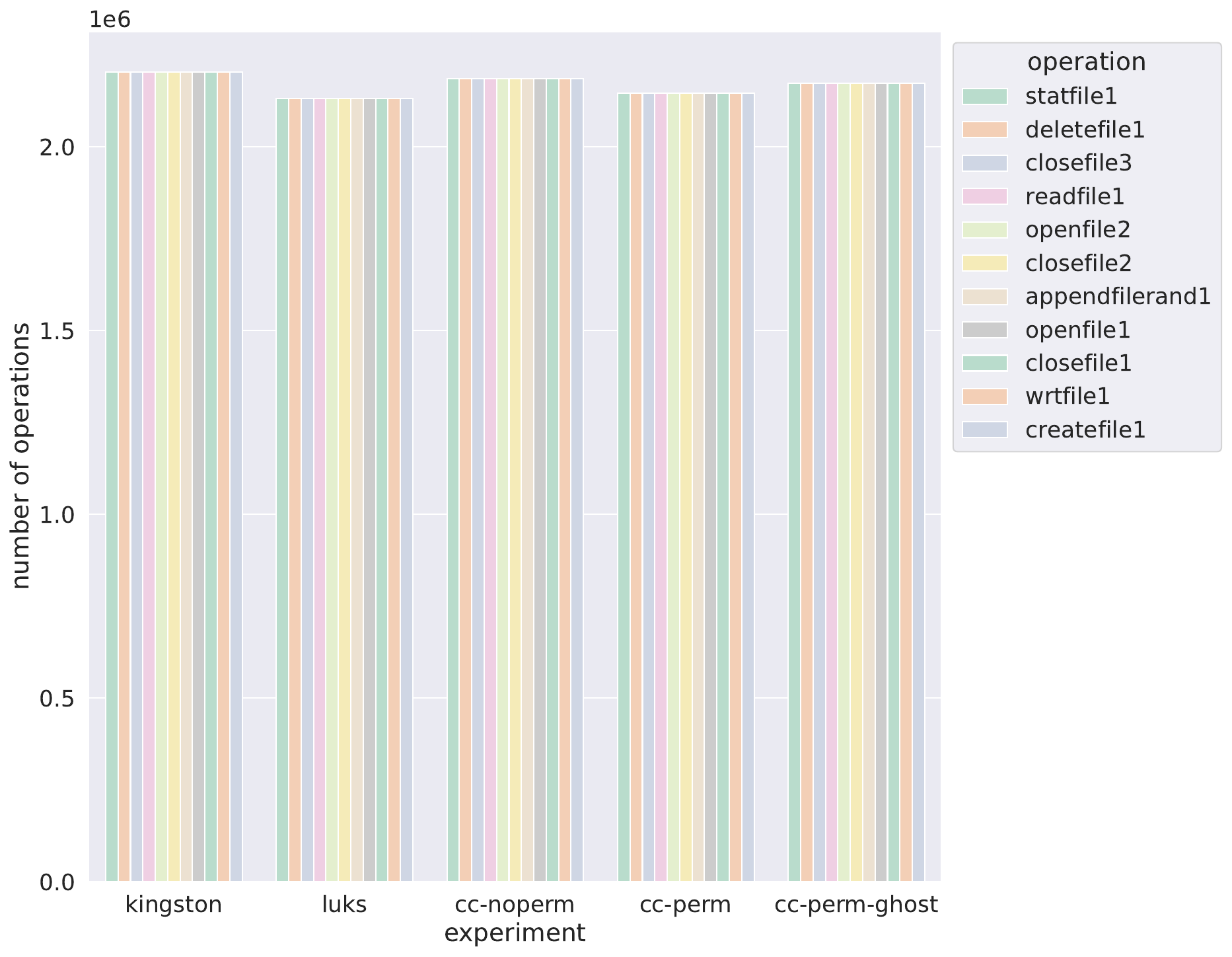}
	}
	\caption{Number of operations completed (by the thousands,
	itemized by the type of operation)
 	for the \texttt{fileserver}
	workload.
	 \label{fileserver-ops}}
\end{center}
\end{figure}

\begin{figure}[ht]
\begin{center}
	\resizebox{0.4\columnwidth}{!}{
		\includegraphics{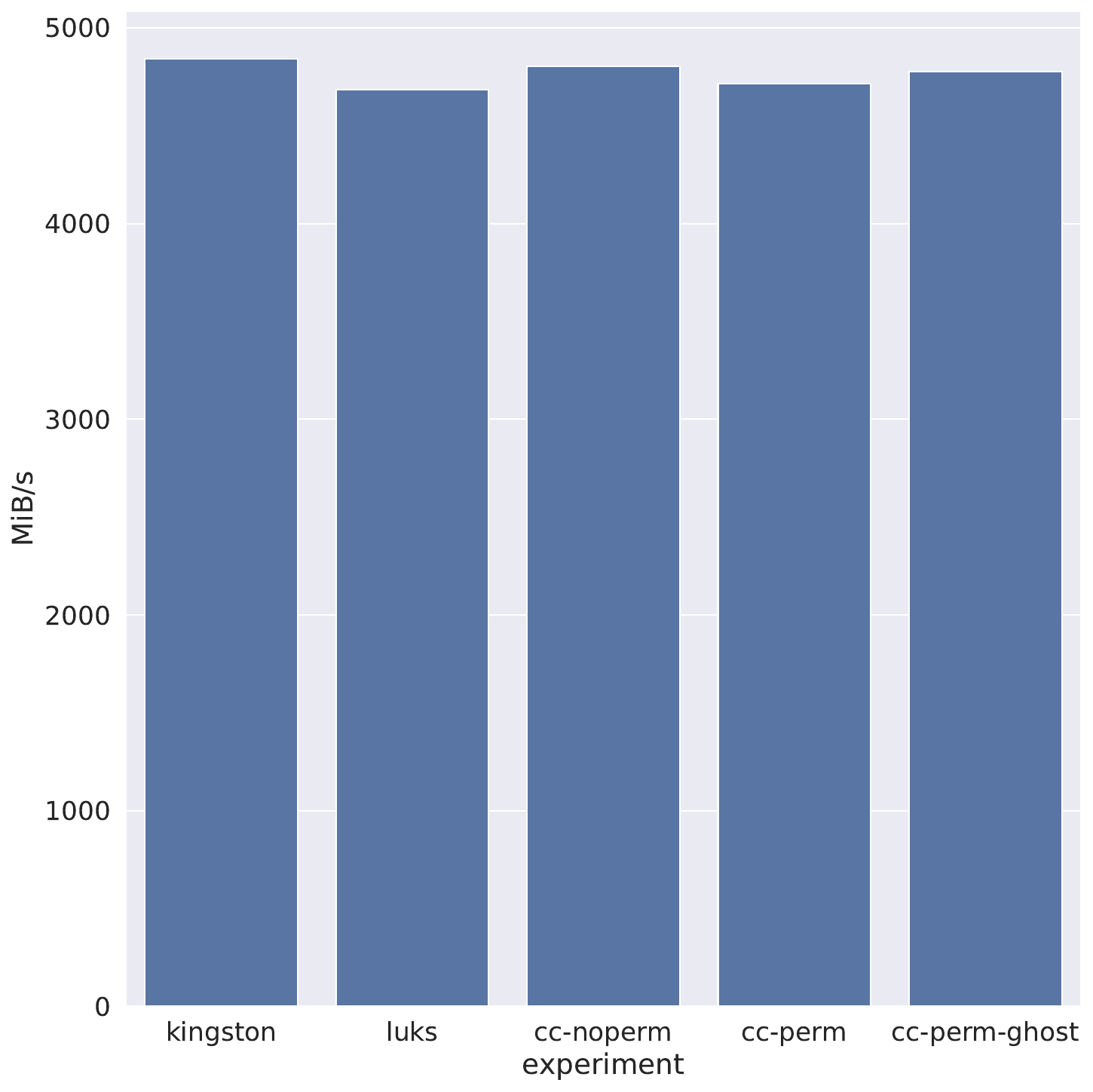}
	}
	\resizebox{0.4\columnwidth}{!}{
		\includegraphics{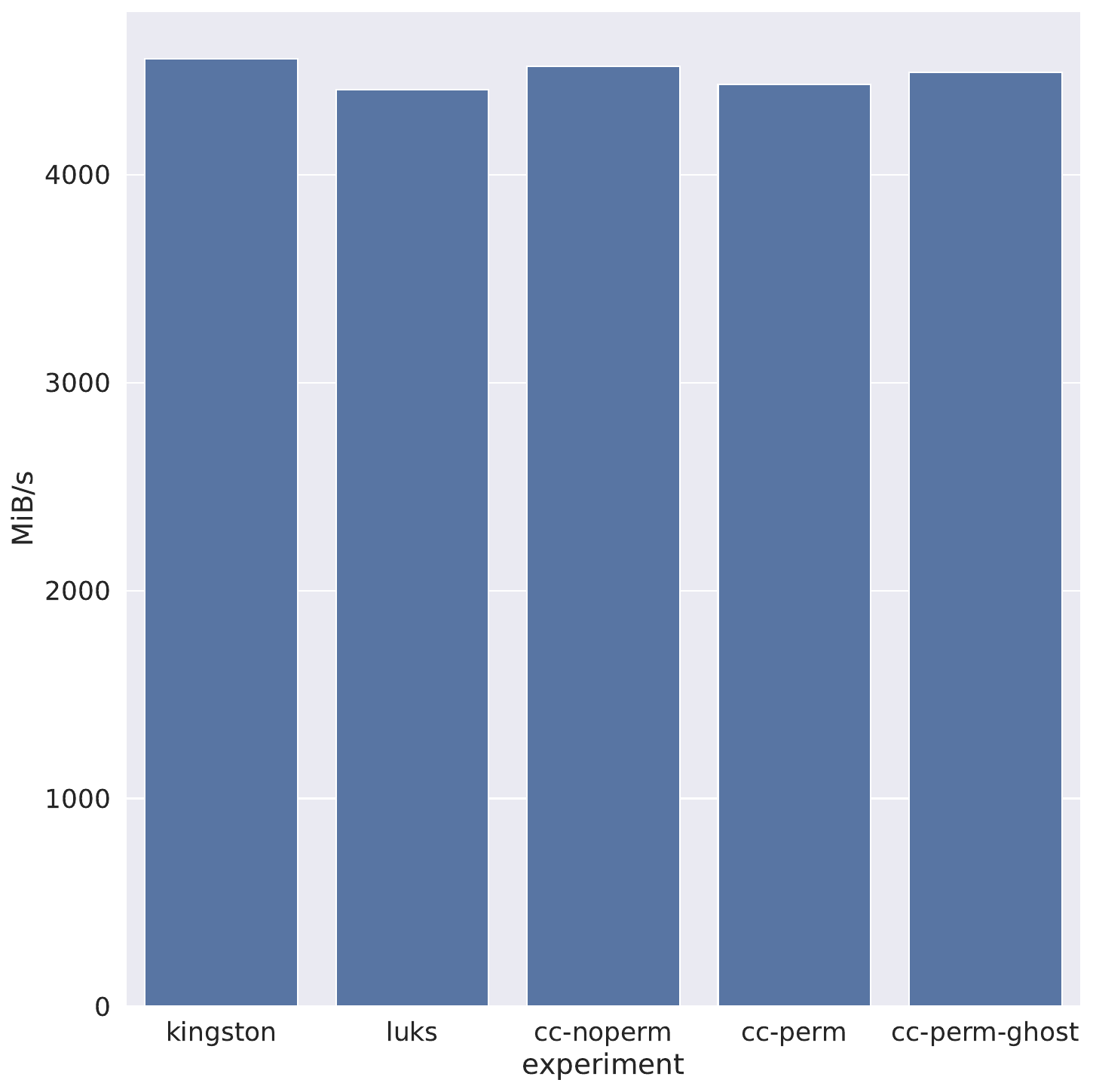}
	}
	\caption{Bandwidth for read and write operations, for the \texttt{fileserver}
	workload.
	 \label{fileserver-bw}}
\end{center}
\end{figure}

In both cases, the host's cache is enabled (the common case when using a disk)
and the workload is executed for 60 seconds.
Figure~\ref{fileserver-ops} shows the number of operations
performed by the first workload
In this case, the set of operations is small and the number of operations are
itemized in the graph.
Figure~\ref{fileserver-bw} shows the bandwidth for read and write operations.

\begin{figure}[ht]
\begin{center}
	\resizebox{0.6\columnwidth}{!}{
		\includegraphics{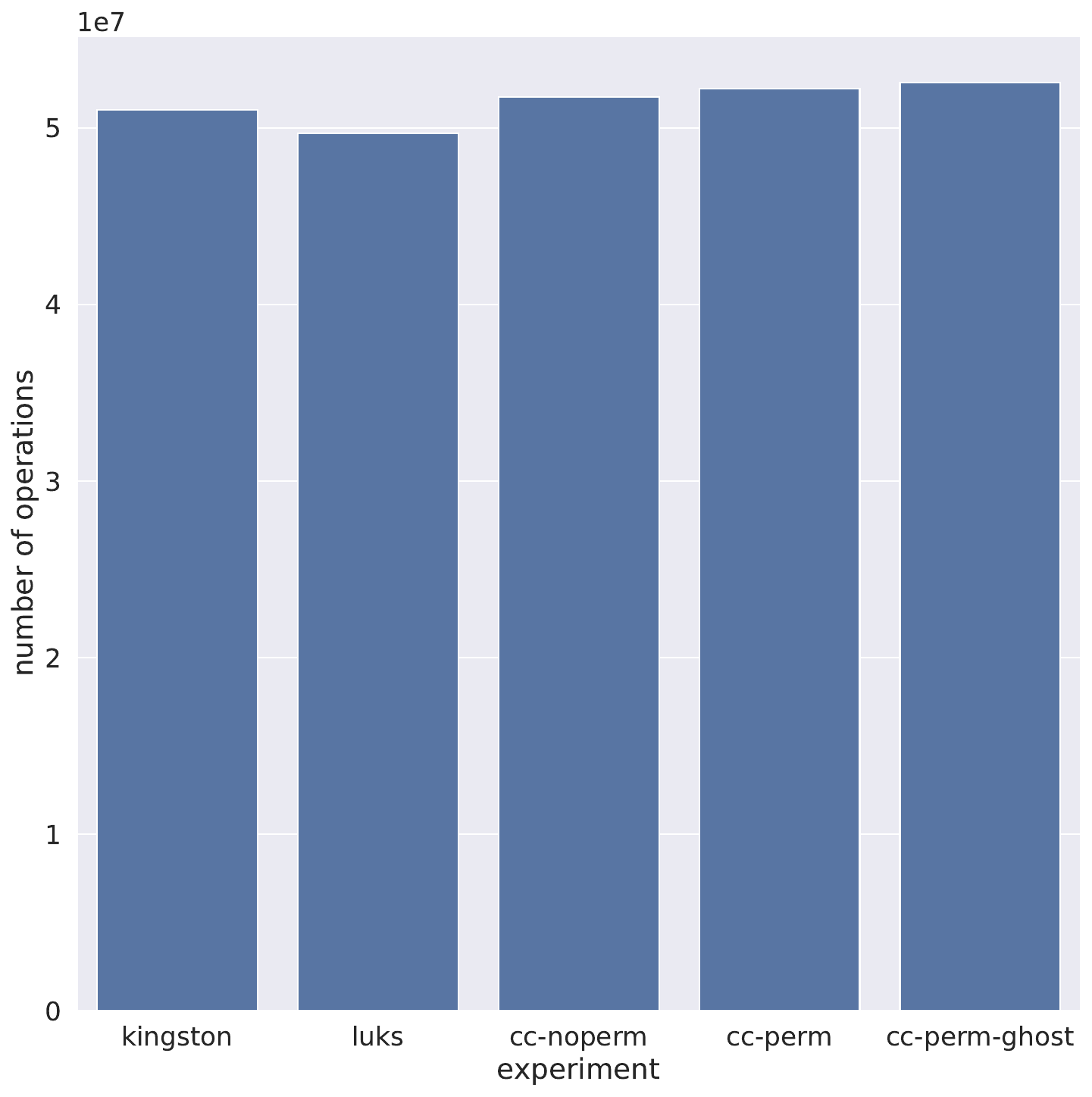}
	}
	\caption{Aggregated number of operations  for the \texttt{webserver}
	workload.
	 \label{webserver-ops}}
\end{center}
\end{figure}

\begin{figure}[ht]
\begin{center}

	\resizebox{0.4\columnwidth}{!}{
		\includegraphics{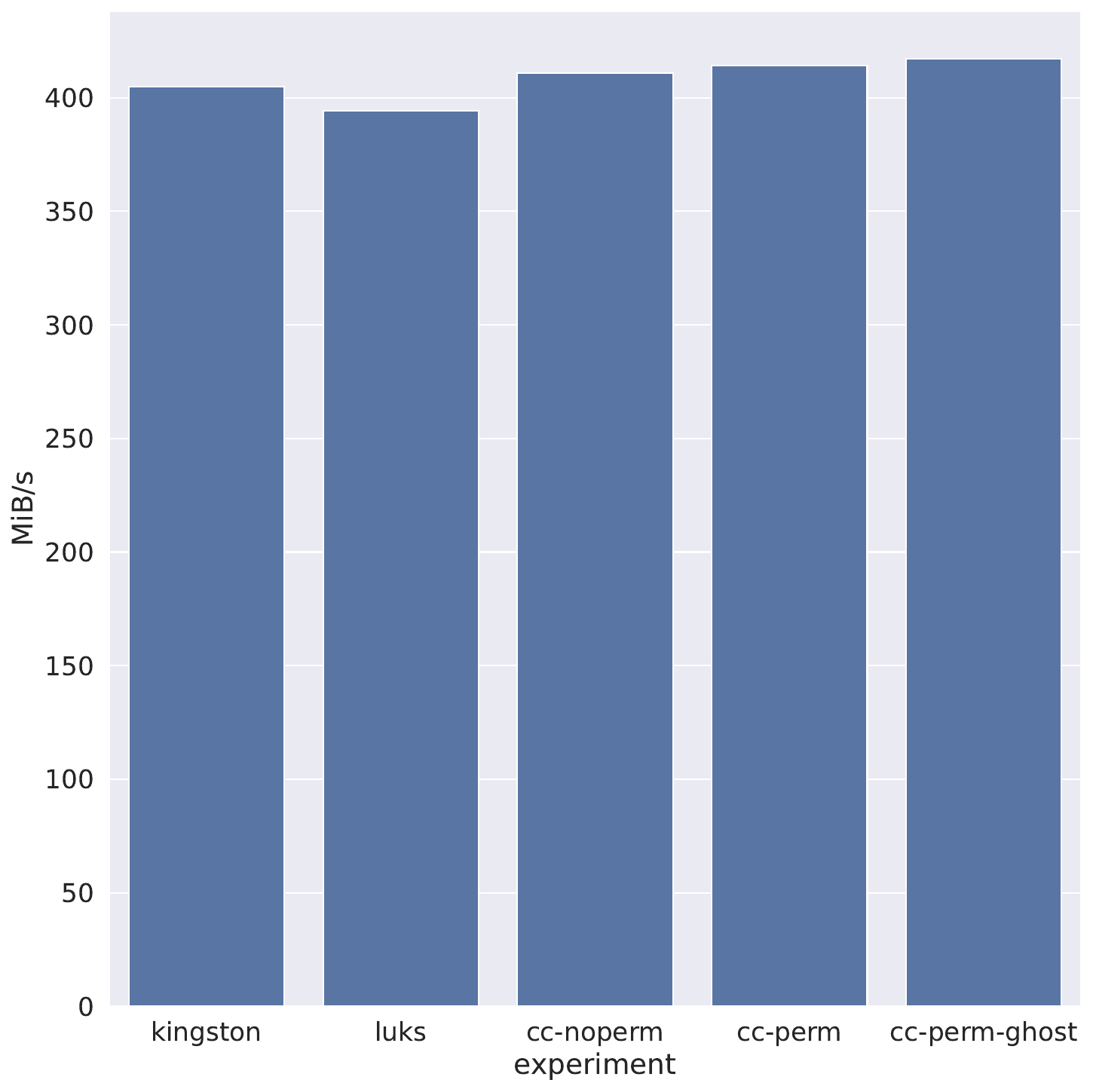}
	}
	\resizebox{0.4\columnwidth}{!}{
		\includegraphics{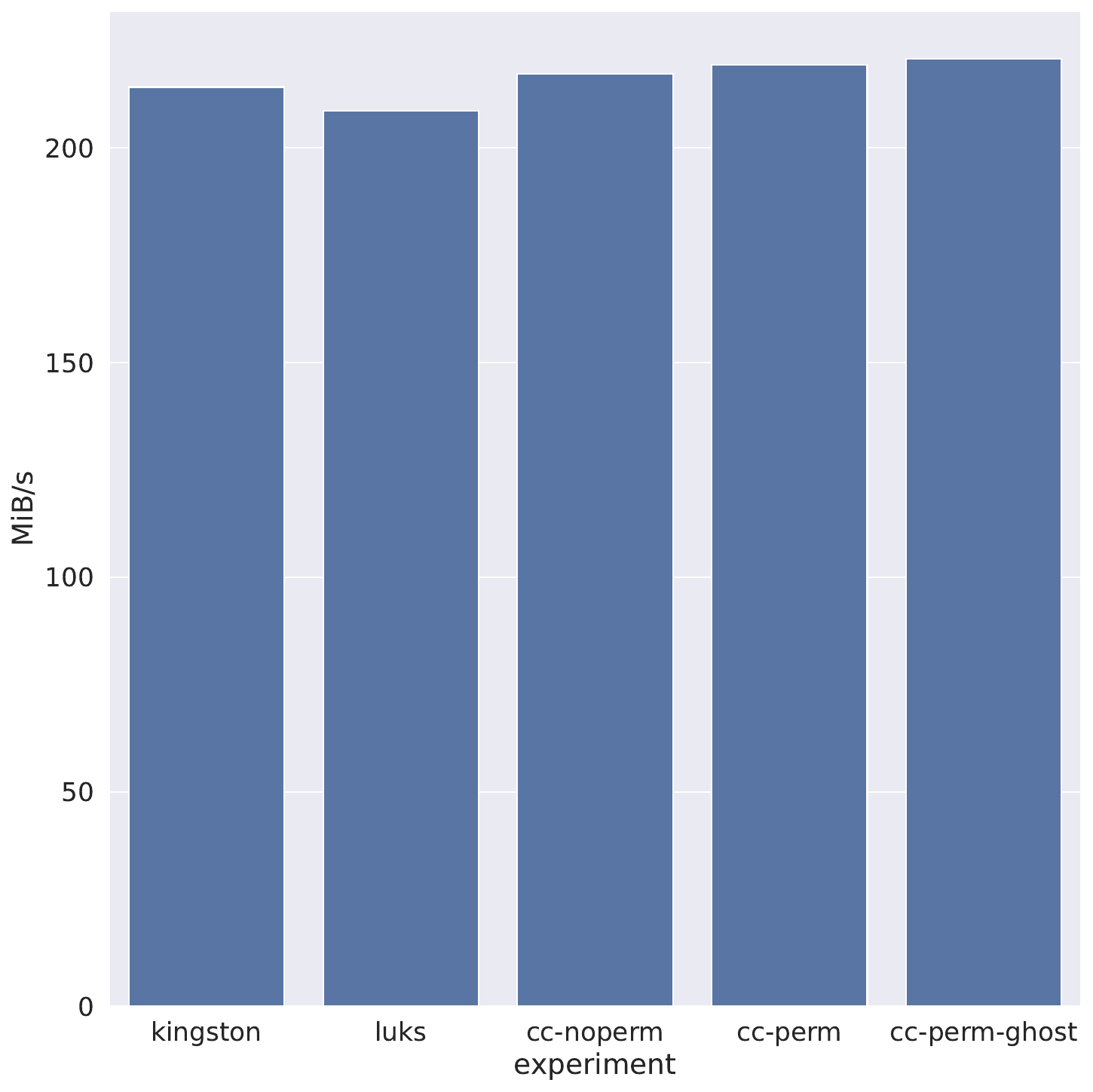}
	}
	\caption{Bandwidth for read (left) and append (right)
	operations, for the \texttt{webserver}
	workload.
	 \label{webserver-bw}}
\end{center}
\end{figure}

Figure~\ref{webserver-ops} shows the results for the second workload.
In this case, the set of operations is bigger and the operations are not itemized,
they are aggregated in one bar.
Figure~\ref{webserver-bw} shows the bandwidth for read and append operations.

\subsection{Experimental Conclusions}
Even though the first results suggest that the performance of the
prototype would not be good enough for practical usage, this is not the
case as shown by the Filebench measurements.

In any case, we can draw four basic conclusions from FIO results with
direct I/O (\texttt{O\_DIRECT}). First, performing direct I/O, the CC
prototype cannot compete with highly optimized standard FDE solutions
such as LUKS in more demanding storage scenarios.

Second, introducing the SBC as an intermediate point has a significant
negative impact on performance, mainly due to the overhead associated with
OTG mechanisms and the additional USB hop required to reach the storage
device.  In most cases, the cost of the CC logic is negligible compared
to the costs of the interposition mechanisms.  In the most favorable
configuration for the CC (i.e. without permutation and with the built-in
cache enabled in the case of sequential read and write operations), the
performance is nearly identical to that of the external drive directly
exported over OTG.  Thus, the overhead introduced by the NBD protocol,
as well as block encryption/decryption and authentication mechanisms,
is negligible compared to the cost of the the USB interposition.
We could further optimize the prototype (e.g. by enabling concurrent
requests, creating a built-in read-ahead cache to mitigate the cost of the
permutation, and selecting more efficient encryption and authentication
algorithms) but these FIO results show that the main bottleneck is imposed
by the interposition mechanisms.
%%V2
Therefore, initial optimization efforts
should primarily focus on enhancing the efficiency of the OTG subsystem
and optimizing the I/O path of the CC, by using mechanisms like
\texttt{io\_uring} or following other approaches
(for example, see ~\cite{3578353.3589545,3650697.3650704,MacedoTHC0022,3805485.3805520}).

Third, the cost of the permutation is very high and affects mainly the
sequential read operations, as expected. Note that systems are optimized
for locality efficiency, from the operating system’s I/O scheduler
and cache (both on the host and the SBC) to the disk firmware. The
permutation breaks all the assumptions made by those components.  We tried
two optimizations in the CC to improve performance. The first was a
write-through cache with read-ahead implemented within the CC program
itself. The second involved forcing to load the disk blocks into the
SBC's operating system cache by detecting sequential access patterns
(i.e., performing read-ahead from the process to populate the system's
cache).  Neither of these changes resulted in a substantial improvement
in sequential read performance (a performance increase was observed,
but it is not relevant for the comparative results, so the data have
not been included in the graphs).

One of the most notable observations from the results of FIO with the
host's cache enabled is the behavior of LUKS, which achieves better
performance in these experiments than the disk directly connected to
the host.  This can be attributed to the system’s highly aggressive
caching policies for LUKS-encrypted volumes in this setup.

Although initial measurements with FIO using the host system cache suggest
that employing CC in a real-world scenario (i.e., with a conventional
file system) would be significantly slower than direct disk access or
LUKS, the Filebench results show that this is not the case.
The results indicate that both the number of operations
performed and the achieved throughput are comparable across all
configurations. The differences across the five scenarios are minimal
and can be attributed to the transient state of the system (e.g.,
effects on locking). The experiments were repeated multiple times,
consistently yielding similar results.

This outcome is somewhat surprising, as it is not intuitive. Ultimately,
the caches in the host performs effectively, masking the performance penalty
of the underlying storage layer. Even the overhead associated with ghost
reads is effectively masked by the efficient use of the Linux cache for
blocks, inodes, and directory entries.
The only visible difference (from the point of view of the user) observed
in other experiments under realistic file system workloads is that the
time required to \emph{unmount} the file system was slightly higher when
using the CC (as expected, because the bandwidth for \texttt{O\_DIRECT}
write operations is lower, so it takes longer to synchronize the volume
before detaching it).

Therefore, the experiments suggest that the approach is viable in
realistic settings.

%%V2
\section{Limitations \label{limitations}}

As stated in the threat model, this approach assumes that the CC software
and hardware are correct and trustworthy. To satisfy these requirements,
the firmware and hardware must provide mechanisms for establishing a root
of trust (additionally, the CC may employ remote attestation mechanisms).
Nevertheless, our initial prototype has been tested and evaluated on
hardware that currently does not provide a root of trust to guarantee
the integrity and authenticity of the system components, for example,
through Secure Boot or hardware-based security mechanisms such as a TPM.
In any case, it should be noted that such mechanisms would not
affect the performance of the CC when providing FADE to the host,
as they would only affect the CC boot time. Therefore, the evaluation
presented in the paper remains valid for assessing the feasibility
of the proposed approach.

\section{Conclusions \label{conclusions}}

This paper has presented the architecture and an implementation of
an alternative approach to Full Disk Encryption (FDE) based on an external
cryptographic device implemented using a general-purpose single-board
computer (SBC) which is located between the host system and a common disk.
By interposing such a programmable USB device between the host and
the storage medium, our design decouples cryptographic functionality from
the operating system, providing a transparent, auditable,
and flexible solution for \emph{at-rest} encryption.

By decoupling encrypted data from the metadata required for encryption,
the resulting storage devices become fully disposable, as they contain
exclusively ciphertext. This significantly increases the difficulty of
cryptanalytic attacks compared to conventional FDE systems. Furthermore,
this approach enables (i) the use of standard encryption modes, in contrast
to those specifically tailored for FDE, which are known to exhibit
additional constraints and limitations; and (ii) to deliver FADE
(Fully Authenticated Disk Encryption) in a simple, straightforward,
and efficient manner.

The proposed architecture offers other notable advantages. It preserves
full transparency for the host system, requiring no
modifications to existing operating systems. Moreover,
it leverages commodity hardware and free and open-source software,
resulting in a low-cost and easily reproducible solution. In addition,
the use of a programmable platform enables rapid adaptation to new
cryptographic algorithms and policies.
Additionally, the interposed encryption model enables CC to implement
countermeasures against untrusted disk firmware, including block
remapping and the generation of fake read operations (i.e. \emph{ghost reads})
to hinder the detection of known-plaintext regions.

We presented a complete experimental evaluation of our research prototype
by using two standard I/O benchmarks. The results show that
the approach is practical and sufficiently efficient for common
real-world scenarios. Although the introduction of an intermediate
device inevitably incurs notable I/O performance overhead,
the different caching layers in both operating systems (host and CC),
as well as within the CC software itself, mask the performance penalty.
The resulting performance is comparable to that of a conventional hard disk.

Nevertheless, there remains significant room for improvement.
The experimental results indicate that initial efforts should focus on
optimizing Linux OTG capabilities.
In addition, the prototype itself can also be substantially enhanced,
for instance by adopting more efficient cryptographic algorithms and
increasing concurrency in the implementation.

The CC is released as free/libre software. It is available in the
following repository:

\begin{center}

 \texttt{\url{https://gitlab.eif.urjc.es/publications/cc}}

\end{center}

\section*{Funding}

This work has been funded in part by the Community of Madrid within the 2023-2026
agreement with  Univ. Rey Juan Carlos for the granting of direct subsidies
for the promotion and encouragement of research and technology transfer,
Line of Action A Emerging Doctors, under Project OrdeNGN (Ref. F1177).

\section*{Acknowledgments}

Generative AI software tools (Microsoft
Copilot\footnote{https://www.bing.com/chat}) have been used exclusively
to edit and improve the quality of human-generated existing text.

\bibliographystyle{IEEEtran}
\bibliography{cc}

\end{document}